\documentclass[a4paper,fleqn,usenatbib]{mnras}
\usepackage{etoolbox}
\makeatletter
\patchcmd\@combinedblfloats{\box\@outputbox}{\unvbox\@outputbox}{}{%
   \errmessage{\noexpand\@combinedblfloats could not be patched}%
}%
 \makeatother
 
\usepackage[T1]{fontenc}
\usepackage{ae,aecompl}

\usepackage{booktabs,caption,fixltx2e}
\usepackage[flushleft]{threeparttable}

\usepackage{graphicx}	
\usepackage{amsmath}	
\usepackage{amssymb}	

\title[Universe opacity and CMB]{Universe opacity and CMB}

\author[V. Vavry\v{c}uk]{
V\'{a}clav Vavry\v{c}uk,\thanks{E-mail: vv@ig.cas.cz}
\\
Institute of Geophysics, The Czech Academy of Sciences, Bo\v{c}n\'{i} II, Praha 4, 14100, Czech Republic\\
}

\date{Accepted 2018 April 16. Received 2018 April 16; in original form 2017 August 15}

\pubyear{2017}

\begin{document}
\label{firstpage}
\pagerange{\pageref{firstpage}--\pageref{lastpage}}
\maketitle%
\begin{abstract}
%
A cosmological model, in which the cosmic microwave background (CMB) is a thermal radiation of intergalactic dust instead of a relic radiation of the Big Bang, is revived and revisited. The model suggests that a virtually transparent local Universe becomes considerably opaque at redshifts $z > 2-3$. Such opacity is hardly to be detected in the Type Ia supernova data, but confirmed using quasar data. The opacity steeply increases with redshift because of a high proper density of intergalactic dust in the previous epochs. The temperature of intergalactic dust increases as $(1+z)$ and exactly compensates the change of wavelengths due to redshift, so that the dust radiation looks apparently like the radiation of the blackbody with a single temperature. The predicted dust temperature is $T^{D} = 2.776 \, \mathrm{K}$, which differs from the CMB temperature by 1.9\% only, and the predicted ratio between the total CMB and EBL intensities is 13.4 which is close to 12.5 obtained from observations. The CMB temperature fluctuations are caused by EBL fluctuations produced by galaxy clusters and voids in the Universe. The polarization anomalies of the CMB correlated with temperature anisotropies are caused by the polarized thermal emission of needle-shaped conducting dust grains aligned by large-scale magnetic fields around clusters and voids. A strong decline of the luminosity density for $z > 4$ is interpreted as the result of high opacity of the Universe rather than of a decline of the global stellar mass density at high redshifts.
\end{abstract}
 
\section{Introduction}

The cosmic microwave background (CMB) radiation was discovered by \citet{Penzias1965} who reported an isotropic and unpolarized signal in the microwave band characterized by a temperature of about 3.5 K. After this discovery, many experiments have been conducted to provide more accurate CMB observations. The rocket measurements of \citet{Gush1990} and  FIRAS on the COBE satellite \citep{Mather1990, Fixsen1996} proved that the CMB has almost a perfect thermal blackbody spectrum with an average temperature of $T = 2.728 \pm 0.004\, \mathrm{K}$ \citep{Fixsen1996}, which was further improved using the WMAP data to $T = 2.72548 \pm 0.00057 \, \mathrm{K}$ \citep{Fixsen2009}.

The CMB temperature consists of small directionally dependent large- and small-scale fluctuations analysed, for example, by the WMAP \citep{Bennett2003} or Planck \citep{Ade2014a,Ade2014b,Aghanim2016} data. The large-scale fluctuation of $\pm0.00335 \,\mathrm{K}$ with one hot pole and one cold pole is called the 'dipole anisotropy' being caused by motion of the Milky Way relative to the Universe \citep{Kogut1993}. The small-scale fluctuations are of about $\pm 70 \,\mu\mathrm{K}$ being studied, for example, by the WMAP \citep{Bennett2003, Hinshaw2009, Bennett2013}, ACBAR \citep{Reichardt2009}, BOOMERANG \citep{MacTavish2006}, CBI \citep{Readhead2004}, and VSA \citep{Dickinson2004} instruments using angular multipole moments. The measurements of the CMB temperature were supplemented by detection of the CMB polarization anomalies by the DASI telescope at a subdegree angular scale by \citet{Kovac2002} and \citet{Leitch2002}. The DASI polarization measurements were confirmed and extended by the CBI \citep{Readhead2004}, CAPMAP \citep{Barkats2005}, BOOMERANG \citep{Montroy2006} and WMAP \citep{Page2007} observations. The measurements indicate that the polarization anomalies and the temperature anisotropies are well correlated.

Immediately, after the discovery of the CMB by \citet{Penzias1965}, \citet{Dicke1965} proposed to interpret the CMB as a blackbody radiation originated in the hot Big Bang. Since the blackbody radiation has been predicted for the expanding universe by several physicists and cosmologists before the CMB discovery \citep{Alpher1948, Gamow1952, Gamow1956}, the detection of the CMB by \citet{Penzias1965} was a strong impulse for a further development of the Big Bang theory. Over years, the CMB radiation became one of the most important evidences supporting this theory. The temperature fluctuations have several peaks attributed to some cosmological parameters such as the curvature of the universe or the dark-matter density \citep{Hu_Dodelson2002, Spergel2007, Komatsu2011}. The polarization anomalies are interpreted as a signal generated by Thomson scattering of a local quadrupolar radiation pattern by free electrons during the last scattering epoch \citep{Hu_White1997}.

The CMB as a relic radiation of the hot Big Bang is now commonly accepted even though it is not the only theory offering an explanation of the CMB origin. Another option, discussed mostly in the cold Big Bang theory \citep{Hoyle1967, Layzer1973, Rees1978, Hawkins1988, Aguirre2000} and in steady-state and quasi-steady-state cosmology \citep{Bondi_Gold1948, Hoyle1948, Arp1990, Hoyle1993, Hoyle1994} is to assume that the CMB does not originate in the Big Bang but is a radiation of intergalactic dust thermalized by the light of stars \citep{Wright1982, Pan1988, Bond1991, Peebles1993, Narlikar2003}. The 'dust theory' assumes the CMB to be produced by dust thermalization at  high redshifts. It needs the high-redshift Universe to be significantly opaque at optical wavelengths which is now supported by observations of the intergalactic opacity \citep{Menard2010a, Xie2015} and by the presence of dust in damped Lyman $\alpha$ absorbers in intergalactic space at high redshift \citep{Vladilo2006, Noterdaeme2017, Ma2017}.

Both the hot Big Bang theory and the dust theory are faced with difficulties when modelling properties of the CMB. The hot Big Bang works well under the assumption of a transparent Universe but it cannot satisfactorily explain how the CMB could survive the opaque epochs of the Universe without being significantly distorted by dust. The distortion should be well above the sensitivity of the COBE/FIRAS, WMAP or Planck flux measurements and should include a decline of the spectral as well as total CMB intensity due to absorption \citep{Vavrycuk2017b}. Detailed analyses of the CMB anisotropies by WMAP and Planck also revealed several unexpected features at large angular scales such as non-Gaussianity of the CMB \citep{Vielva2004, Cruz2005, Planck_XXIV_2014} and a violation of statistical isotropy and scale invariance following from the Big Bang theory \citep{Schwarz2016}.

By contrast, the dust theory has troubles with explaining why intergalactic dust radiates as a blackbody, why the CMB temperature is unaffected by a variety of redshifts of radiating dust grains \citep[p. 289]{Peacock1999}, and why the CMB is almost isotropic despite the variations of the dust density and of the starlight in the Universe. The theory should also satisfactorily explain correlated observations of the temperature and polarization fluctuations in the CMB.

The assumption of the blackbody radiation of intergalactic dust was questioned, for example, by \citet{Purcell1969} who analysed the Kramer-Kronig relations applied to space sparsely populated by spheroidal grains. He argued that intergalactic dust grains whose size is less than 1 $\mu$m are very poor radiators of millimetre waves and thus cannot be black. On the other hand, \citet{Wright1982} demonstrated that needle-shaped conducting grains could provide a sufficient long-wavelength opacity. The long-wavelength absorption is also strengthened by complex fractal or fluffy dust aggregates \citep{Wright1987, Henning1995}. Hence, it now seems that the opacity of intergalactic dust is almost unconstrained and the assumption of the blackbody radiation of intergalactic dust is reasonable \citep{Aguirre2000}.

In this paper, I address the other objections raised to the dust theory. I show that under some assumptions about the stellar and dust mass evolution in the Universe the idea of the CMB produced by dust thermalization can be reconciled with observations and that the controversies of the dust theory might be apparent. I present formulas for the redshift-dependent extragalactic background light (EBL), which is the main source of the intergalactic dust radiation. Subsequently, I determine the redshift-dependent temperature of dust and establish a relation between the intensity of the EBL and CMB. Based on observations of the opacity of the Universe, the maximum redshift of dust contributing to the observed CMB is estimated. Finally, I discuss why the CMB temperature is so stable and how the small-scale temperature and polarization anisotropies in the CMB can be explained in the dust theory. 

\section{Extragalactic background light (EBL)}

\subsection{Observations of EBL}

The EBL covers the near-ultraviolet, visible and infrared wavelengths from 0.1 to 1000 $\mu$m and has been measured by direct as well as indirect techniques. The direct measurements were provided, for example, by the IRAS, DIRBE on COBE, FIRAS, ISO, and SCUBA instruments, for reviews see \citet{Hauser2001, Lagache2005, Cooray2016}. The direct measurements are supplemented by analysing integrated light from extragalactic source counts \citep{Madau2000, Hauser2001} and attenuation of gamma rays from distant blazars due to scattering on the EBL \citep{Kneiske2004, Dwek2005, Primack2011, Gilmore2012}.
 
\begin{figure*}
\includegraphics[angle=0,width=13cm,trim=60 80 60 60, clip=true]{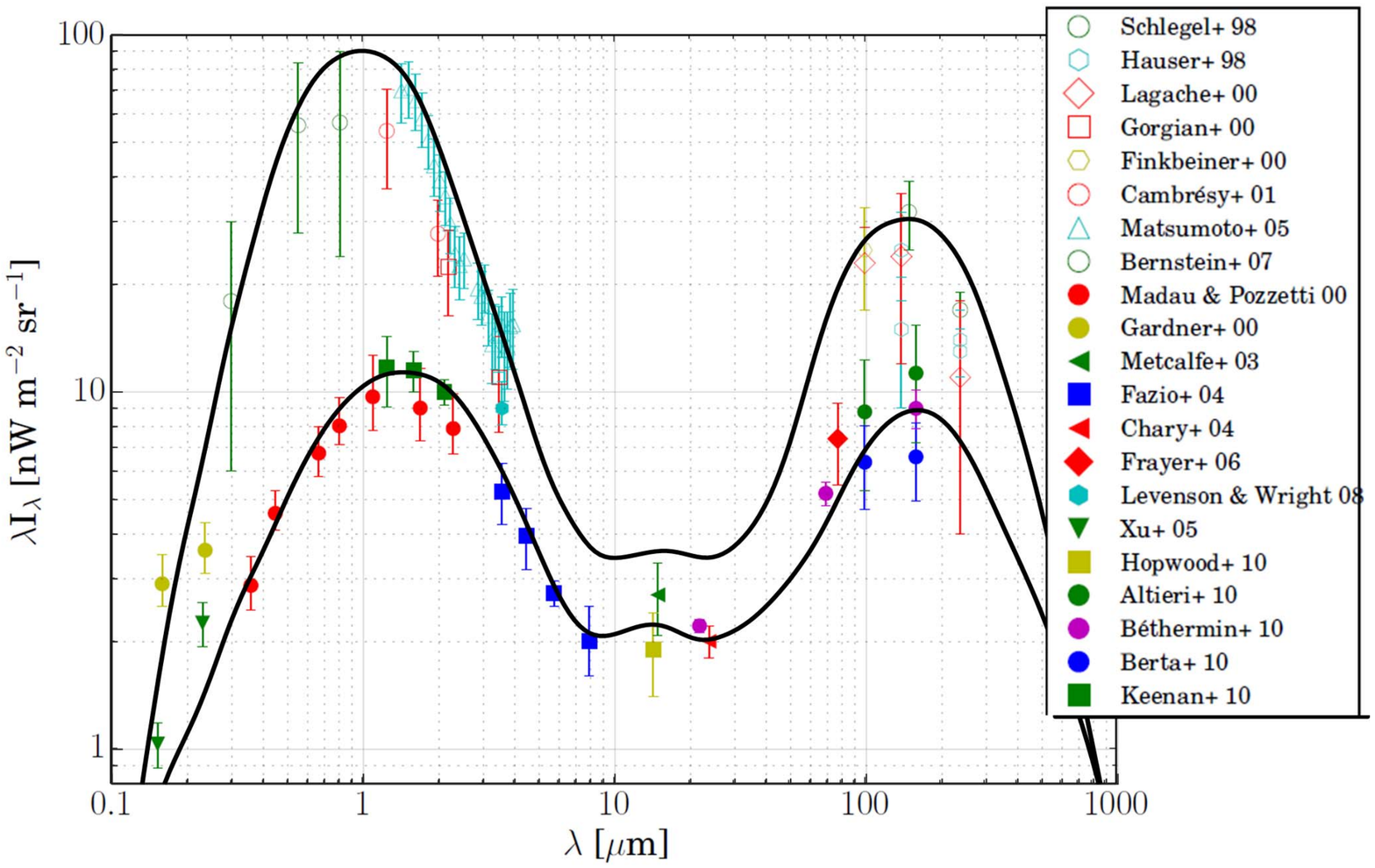}
\caption{
Spectral energy distribution (SED) of the extragalactic background light (EBL) with estimates of its minimum and maximum limits (black lines). The observations reported by various authors are marked by colour symbols (modified after \citet{Dominguez2011}).
}
\label{fig:1}
\end{figure*}

The spectrum of the EBL has two distinct maxima: at visible-to-near-infrared wavelengths (0.7 - 2 $\mu$m) and at far-infrared wavelengths (100-200 $\mu$m) associated with the radiation of stars and with the thermal radiation of dust in galaxies \citep{Schlegel1998, Calzetti2000}. Despite the extensive measurements of the EBL, the uncertainties are still large (see Figure~\ref{fig:1}). The best constraints at the near- and mid-infrared wavelengths come from the lower limits based on the integrated counts \citep{Fazio2004, Dole2006, Thompson2007} and from the upper limits based on the attenuation of gamma-ray photons from distant extragalactic sources \citep{Dwek2005, Aharonian2006, Stecker2006, Abdo2010}. Integrating the lower and upper limits of the spectral energy distributions shown in Figure~\ref{fig:1}, the total EBL should fall between 40 and $200 \,\mathrm{n W m}^{-2}\mathrm{sr}^{-1}$. The most likely value of the total EBL from 0.1 to 1000 $\mu$m is about $80-100 \, \mathrm{n W m}^{-2}\mathrm{sr}^{-1}$ \citep{Hauser2001}.

\subsection{Redshift dependence of EBL}

Let us assume that the comoving galaxy and dust number densities, galaxy luminosity and galactic and intergalactic opacities are conserved with cosmic time. The EBL in a transparent expanding universe with no light sources is calculated by the equation of radiative transfer \citep[his equation 5.158 with luminosity density $j=0$]{Peebles1993}
\begin{equation}\label{eq1}
\frac{d}{dt}I_\nu^{\mathrm{EBL}} +3HI_\nu^{\mathrm{EBL}} = 0 \,,
\end{equation}
where $I_\nu^{\mathrm{EBL}}(t)$  is the specific intensity of the EBL (in $\mathrm{Wm}^{-2}\mathrm{Hz}^{-1}\mathrm{sr}^{-1}$), $H(t)=\dot{R}/R$ is the Hubble parameter and $R$ is the scale factor. Solving equation (1)
\begin{equation}\label{eq2}
\int \frac{dI_\nu^{\mathrm{EBL}}}{I_\nu^{\mathrm{EBL}}} = \int -3Hdt \,,
\end{equation}
and taking into account the time-redshift relation \citep[his equation 13.40]{Peebles1993} 
\begin{equation}\label{eq3}
dt = \frac{1}{H}\frac{dz}{1+z}\,,
\end{equation}
the specific intensity $I_\nu^{\mathrm{EBL}}$ at redshift $z$ is
\begin{equation}\label{eq4}
I_\nu^{\mathrm{EBL}} = \left(1+z\right)^3 I_{\nu0}^{\mathrm{EBL}} \,,
\end{equation}
and subsequently the bolometric intensity $I^{\mathrm{EBL}}$ at redshift $z$
\begin{equation}\label{eq5}
I^{\mathrm{EBL}}\left(z\right) = \left(1+z\right)^4  I_0^{\mathrm{EBL}}  \, ,
\end{equation}
where $I_{\nu0}^{\mathrm{EBL}}$ and $I_{0}^{\mathrm{EBL}}$ are the specific and bolometric EBL intensities at $z=0$. Equation (5) expresses the fact that the bolometric EBL scales with the expansion as $(1+z)^{-4}$. Since light sources and absorption are not considered, the EBL declines with the expansion of the Universe. The decline is $(1+z)^{-3}$ due to the volume expansion and $(1+z)^{-1}$ due to the redshift.  

If light sources and absorption are considered, the equation of radiative transfer must be modified \citep[his equation 12.13]{Peacock1999}
\begin{equation}\label{eq6}
\frac{d}{dt} I_\nu^{\mathrm{EBL}} + 3HI_\nu^{\mathrm{EBL}} = \frac{c}{4\pi}j_\nu - c\kappa_\nu I_\nu^{\mathrm{EBL}} \,,
\end{equation}
where $j_\nu(t)$ is the luminosity density at frequency $\nu$ (in $\mathrm{Wm}^{-3}\mathrm{Hz}^{-1}$) and $\kappa_\nu$ is the opacity at frequency $\nu$. 

If we assume that the comoving stellar and dust mass densities are constant, the comoving specific intensity of the EBL is also constant. Consequently, the radiation from light sources is exactly compensated by light absorption and the right-hand side of equation (6) is zero. It physically means that the light produced by stars is absorbed by galactic and intergalactic dust. The process is stationary because the energy absorbed by intergalactic dust produces its thermal radiation which keeps the dust temperature constant. Since dust grains are very small, any imbalance between the radiated stellar energy and energy absorbed by dust would produce fast observable changes in the dust temperature. Hence, 
\begin{equation}\label{eq7}
I_\nu^{\mathrm{EBL}} = \frac{1}{4\pi} \frac{j_\nu}{\kappa_\nu} \,,
\end{equation}
which should be valid for all redshifts $z$. The specific luminosity density $j_\nu (z)$ in equation (7) increases with $z$ as $(1+z)^3$ and the opacity $\kappa_\nu$ is redshift independent, because the number of absorbers in the comoving volume is constant (the proper attenuation coefficient per unit ray path increases with $z$ but the proper length of a ray decreases with $z$). Hence equation (7) predicts $I_\nu^{\mathrm{EBL}}$ to increase with $z$ as $(1+z)^3$ similarly as in the case of no light sources and no absorption, see equation (4). Consequently, the bolometric EBL intensity increases with $z$ in an expanding dusty universe with galaxies according to equation (5) derived originally for the transparent universe with no light sources.

\subsection{EBL and the Tolman relation}

Equation (5) can alternatively be derived using the Tolman relation, which expresses the redshift dependence of the surface brightness of galaxies in the expanding universe \citep[his equation 3.90]{Peacock1999}
\begin{equation}\label{eq8}
B^{\mathrm{bol}}\left(z\right) = \left(1+z\right)^{-4}B_0^{\mathrm{bol}}  \,,
\end{equation}
where $B^{\mathrm{bol}}\left(z\right)$ and $ B_0^{\mathrm{bol}}$ is the bolometric surface brightness of a galaxy (in $\mathrm{W m}^{-2} \mathrm{sr}^{-1}$) at redshift $z$ and at $z = 0$, respectively. The Tolman relation says that the bolometric surface brightness of galaxies decreases with the redshift in the expanding universe in contrast to the static universe, where the surface brightness of galaxies is independent of their distance. 

In the Tolman relation, the observer is at $z = 0$ and the redshift dependence is studied for distant galaxies at high redshift. However, the relation can be reformulated for an observer at redshift $z$. Obviously, if we go back in time, the surface brightness of galaxies was higher in the past than at present by factor $(1+z)^4$. If the number density of galaxies is assumed to be constant in the comoving volume, the EBL for an observer at redshift $z$ should also be higher by the same factor, see equation (8). Hence, the intensity of the EBL was significantly higher at redshift $z$ than at present. 

Strictly speaking, the Tolman relation was originally derived for a transparent universe with no dust. The presence of dust reduces the intensity of the EBL and dust must be incorporated into the model. In analogy to the surface brightness of galaxies, we can introduce a surface absorptivity of dust as a surface brightness of negative value. Thus instead of radiating energy, the energy is absorbed. Since the dust density is conserved in the comoving volume similarly as the number density of galaxies, the intensity of the EBL will be lower in the partially opaque universe than in the transparent universe, but the redshift dependence described in equation (8) is conserved. 

\section{Opacity observations}

\subsection{Galactic and intergalactic opacity}

The methods for measuring the galactic opacity usually perform multi-wavelength comparisons and a statistical analysis of the colour and number count variations induced by a foreground galaxy onto background sources (for a review, see  \citet{Calzetti2001}). The most transparent galaxies are ellipticals with an effective extinction  $A_V$ of $0.04 - 0.08$ mag. The dust extinction in spiral and irregular galaxies is higher. \citet{Holwerda2005a} found that the dust opacity of the disk in the face-on view apparently arises from two distinct components: an optically thicker component ($A_I = 0.5 - 4$ mag) associated with the spiral arms and a relatively constant optically thinner disk ($A_I = 0.5$ mag). Typical values for the inclination-averaged extinction are:  $0.5 - 0.75$ mag for Sa-Sab galaxies, $0.65 - 0.95$ mag for the Sb-Scd galaxies, and $0.3 - 0.4$ mag for the irregular galaxies at the B-band \citep{Calzetti2001}. 

Adopting estimates of the relative frequency of specific galaxy types in the Universe and their mean visual extinctions \citep[his table 2]{Vavrycuk2017a}, we can estimate their mean visual opacities and finally the overall mean galactic opacity. According to \citet{Vavrycuk2017a}, the average value of the visual opacity $\kappa$ is about $0.22 \pm 0.08$. A more accurate approach should take into account statistical distributions of galaxy sizes and of the mean galaxy surface brightness for individual types of galaxies. 

\citet{Menard2010a} estimated the visual intergalactic attenuation to be $A_V = (1.3 \pm 0.1) \times 10^{-2}$ mag  at a distance from a galaxy of up to 170 kpc and $A_V = (1.3 \pm 0.3) \times 10^{-3}$ mag on a large scale at a distance of up to 1.7 Mpc. Similar values are reported by \citet{Muller2008} and \citet{Chelouche2007} for the visual attenuation produced by intracluster dust. However, the intergalactic attenuation is redshift dependent. It increases with redshift, and a transparent universe becomes significantly opaque (optically thick) at redshifts of $z = 1-3$ \citep{Davies1997}. The increase of intergalactic extinction with redshift is confirmed by \citet{Menard2010a} by correlating the brightness of $\sim$85.000 quasars at $z > 1$ with the position of 24 million galaxies at $z \sim 0.3$ derived from the Sloan Digital Sky Survey. The authors estimated $A_V$ to about 0.03 mag at $z = 0.5$ but to about $0.05 - 0.09$ mag at $z = 1$. In addition, a consistent opacity was reported by \citet{Xie2015} who studied the luminosity and redshifts of the quasar continuum of $\sim 90.000$ objects. The authors estimated the effective dust density $n \sigma_V \sim 0.02 \,h \, \mathrm{Gpc}^{-1}$ at $z < 1.5$. 

Dust extinction can also be estimated from the hydrogen column densities studied by the Lyman $\alpha$ (Ly$\alpha$) absorption lines of damped Lyman absorbers (DLAs). \citet{Bohlin1978} determined the column densities of the interstellar H I towards 100 stars and found a linear relationship between the total hydrogen column density, $N_\mathrm{H} = 2 N_\mathrm{H2} + N_\mathrm{HI}$ , and the colour excess from the Copernicus data
\begin{equation}\label{eq9}
N_\mathrm{H} / \left(A_B-A_V\right) = 5.8 \times 10^{21} \, \mathrm{cm}^{-2} \, \mathrm{mag}^{-1} \,, 
\end{equation}
and
\begin{equation}\label{eq10}
N_\mathrm{H} / A_V \approx 1.87 \times 10^{21} \, \mathrm{cm}^{-2} \, \mathrm{mag}^{-1}\, \, \mathrm{for} \,\, R_V = 3.1\,. 
\end{equation}
\citet{Rachford2002} confirmed this relation using the FUSE data and adjusted slightly the slope in equation (9) to $5.6 \times 10^{21} \, \mathrm{cm}^{-2} \, \mathrm{mag}^{-1}$. Taking into account observations of the mean cross-section density of DLAs reported by \citet{Zwaan2005} 
\begin{equation}\label{eq11}
\langle n \sigma \rangle =  \left(1.13 \pm 0.15 \right) \times 10^{-5} \, h \, \mathrm{Mpc}^{-1} \,,
\end{equation}
the dominating column density of DLAs, $N_\mathrm{HI} \sim 10^{21} \, \mathrm{cm}^{-2}$ \citep{Zwaan2005}, and the mean molecular hydrogen fraction in DLAs of about $0.4-0.6$ \citep[their Table 8]{Rachford2002}, equation (10) yields for the intergalactic attenuation $A_V$ at $z = 0$:  $A_V \sim 1-2 \times 10^{-5} \, h \, \mathrm{Mpc}^{-1}$. Considering also a contribution of less massive LA systems, we get basically the result of \citet{Xie2015}: $A_V \sim 2 \times 10^{-5} \, h \, \mathrm{Mpc}^{-1}$.

\subsection{Wavelength-dependent opacity}

\begin{figure}
\includegraphics[angle=0,width=9.6cm,trim= 90 30 20 40, clip=true]{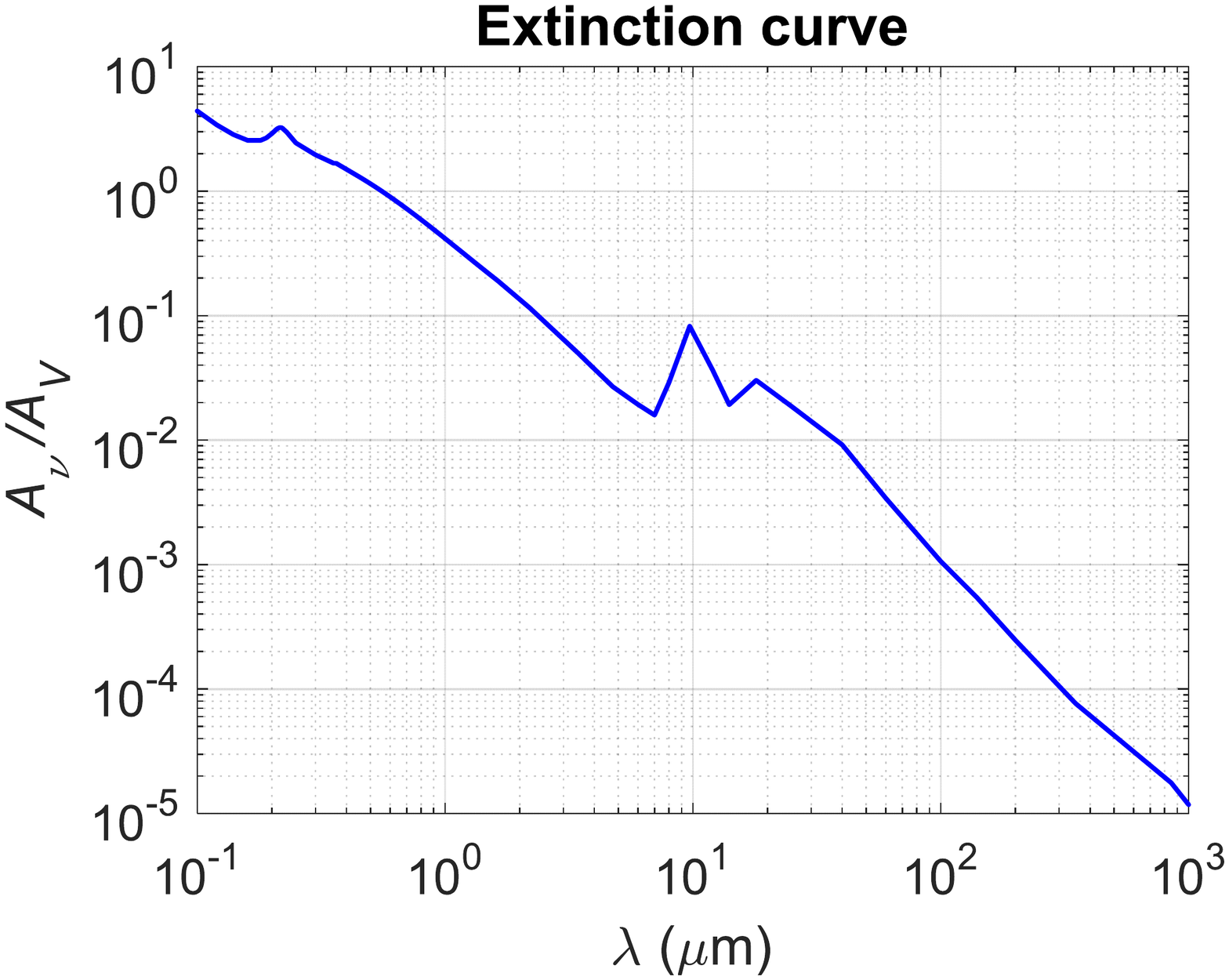}
%
\caption{
Normalized frequency-dependent attenuation due to absorption by dust in the Milky Way \citep[his Tables 4-6]{Draine2003}.
}
\label{fig:2}
\end{figure}

The dust opacity is frequency dependent (see Figure~\ref{fig:2}). In general, it decreases with increasing wavelength but displays irregularities. The extinction law for dust in the Milky Way is well reproduced for infrared wavelengths between $\sim 0.9 \, \mu$m and $\sim 5 \, \mu$m by the power-law $A_\lambda \sim \lambda^{-\beta}$ with $\beta$ varying from 1.3 to 1.8 \citep{Draine2003}. At wavelengths of 9.7 and 18 $\mu$m, the dust absorption displays two maxima associated with silicates \citep{Mathis1990, Li2001, Draine2003}. At longer wavelengths, the extinction decays according to a power-law with $\beta=2$. This decay is predicted by numerical modelling of graphite or silicate dust grains as spheroids with sizes up to 1 $\mu$m \citep{Draine1984}. However, the long-wavelength opacity also depends on the shape of the dust grains. For example, \citet{Wright1982, Henning1995, Stognienko1995} and others report that needle-shaped conducting grains or complex fractal or fluffy dust aggregates can produce much higher long-wavelength opacity than spheroidal grains with the power-law described by $0.6 < \beta < 1.4$ \citep{Wright1987}. 

\subsection{Opacity ratio}

Extinction of the EBL is caused by two effects: (1) the galactic opacity causing obscuration of background galaxies by partially opaque foreground galaxies, and (2) the intergalactic opacity produced by light absorption due to intergalactic dust. The distribution of the absorbed EBL energy between galaxies and intergalactic dust can be quantified by the opacity ratio \citep[his equation 16]{Vavrycuk2017a}
\begin{equation}\label{eq12}
R_\kappa = \frac{\lambda_0 \gamma_0}{\kappa} \,,
\end{equation}
where $\kappa$ is the mean bolometric galactic opacity, $\lambda_0$ is the mean bolometric intergalactic absorption coefficient along a ray path at $z = 0$, and $\gamma_0$ is the mean free path of a light ray between galaxies at $z = 0$, 
\begin{equation}\label{eq13}
\gamma_0 = \frac{1}{n_0 \pi a^2} \,,
\end{equation}
where $a$ is the mean galaxy radius, and $n_0$ is the galaxy number density at $z=0$. 

The opacity ratio is a fundamental cosmological quantity controlled by the relative distribution of dust masses between galaxies and the intergalactic space. Since opacity is a relative quantity, it is invariant to the extinction law and redshift.

\begin{figure}
\includegraphics[angle=0,width=10cm,trim = 180 100 60 80, clip=true]{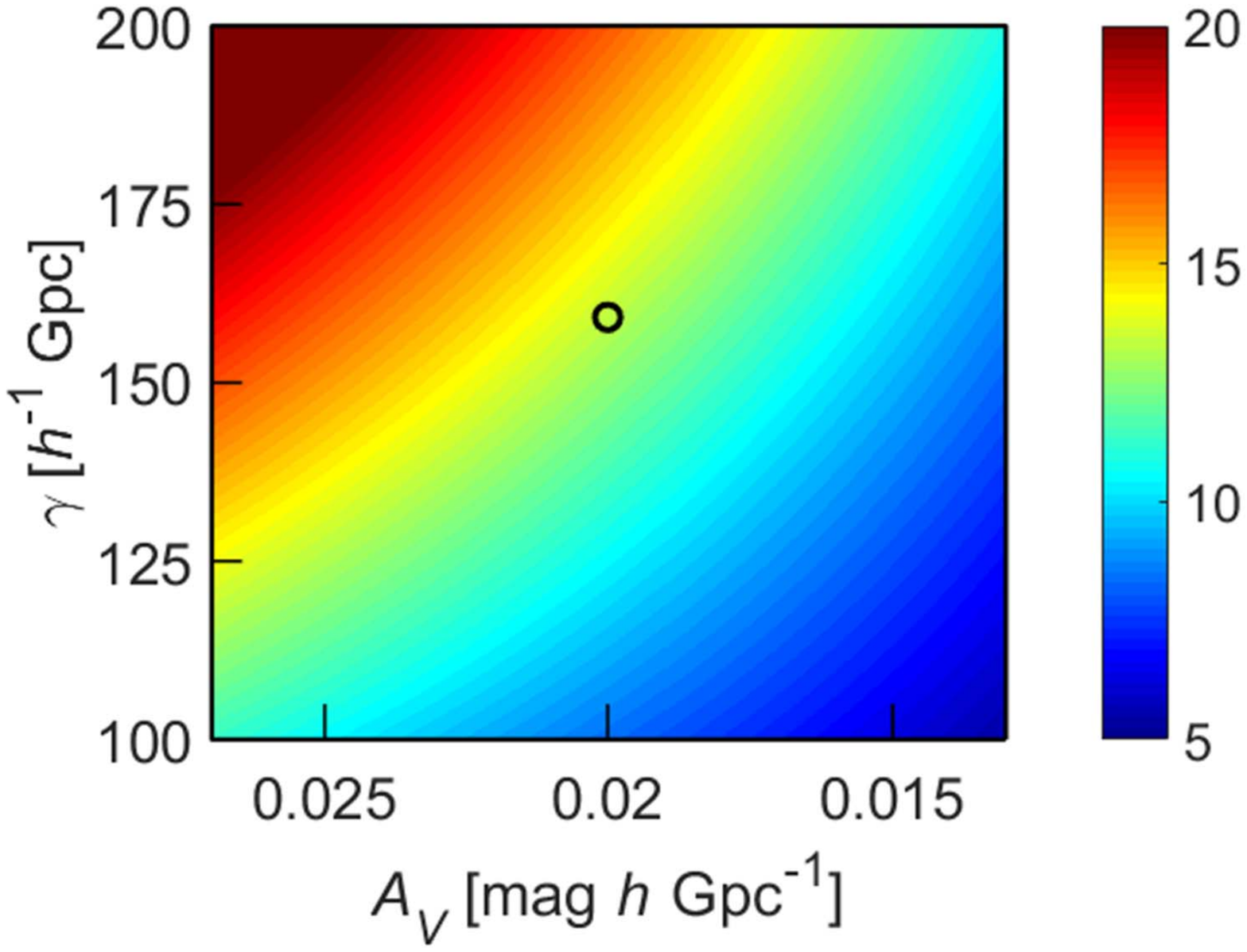}
\caption{
The opacity ratio $R_\kappa$ as a function of the intergalactic visual attenuation $A_V$ and the mean free path between galaxies $\gamma$.
}
\label{fig:3}
\end{figure}

Considering observations of the galactic and intergalactic opacity, and estimates of the mean free path of light between galaxies (see Table~\ref{Table:1}), the opacity ratio is in the range of 6-35 with an optimum value of 13.4 (see Figure~\ref{fig:3}). This indicates that the EBL is predominantly absorbed by intergalactic dust. The EBL energy absorbed by galaxies is much smaller being only a fraction of the EBL energy absorbed by intergalactic dust.

\begin{table*}
%
%
\caption{Opacity ratio}  

\label{Table:1}      
%
%
\centering                          
\begin{tabular}{c c c c c c c c c}  
%
%
\hline\hline                 
%
%
 & $n$ & $\gamma$ & $\kappa$ & $A_V$ & $\lambda_V$ &  $I^{\mathrm{EBL}}$  & $R_\kappa^A$ & $R_\kappa^B$\\
 & $(h^3\,\mathrm{Mpc}^{-3})$ & $(h^{-1}\,\mathrm{Gpc})$ &  & $(\mathrm{mag}\,h\,\mathrm{Gpc}^{-1})$ & $(h\,\mathrm{Gpc}^{-1})$ & $(\mathrm{n W m}^{-2}\,\mathrm{sr}^{-1})$ &  & \\
%
%
\hline                        
%
%
Minimum ratio & 0.025 & 130 & 0.30 & 0.015 & 0.0138 & 200 & 6.0 &  5.0 \\
Maximum ratio & 0.015 & 210 & 0.14 & 0.025 & 0.0230 &  40 & 34.5 & 24.9 \\    
Optimum ratio & 0.020 & 160 & 0.22 & 0.020 & 0.0184 & 80 &  13.4 & 12.5 \\
%
%
\hline                                  
\end{tabular}
%
%
%
\begin{tablenotes}
\item 
$n$ is the number density of galaxies, 
$\gamma$ is the mean free path between galaxies defined in equation (13), 
$A_V$ is the visual intergalactic extinction, 
$\lambda_V$ is the visual intergalactic extinction coefficient, 
$\kappa$ is the mean visual opacity of galaxies, 
$I^{\mathrm{EBL}}$ is the total EBL intensity,
$R_\kappa^A$ is the opacity ratio calculated using equation (12), and 
$R_\kappa^B$ is the opacity ratio calculated using equation (22). 
The mean effective radius of galaxies $a$ is considered to be 10 kpc in equation (13). All quantities are taken at $z = 0$.
\end{tablenotes}
\end{table*}

\section{Thermal radiation of intergalactic dust}

\begin{figure}
\includegraphics[angle=0,width=8cm,trim=100 70 160 70, clip=true]{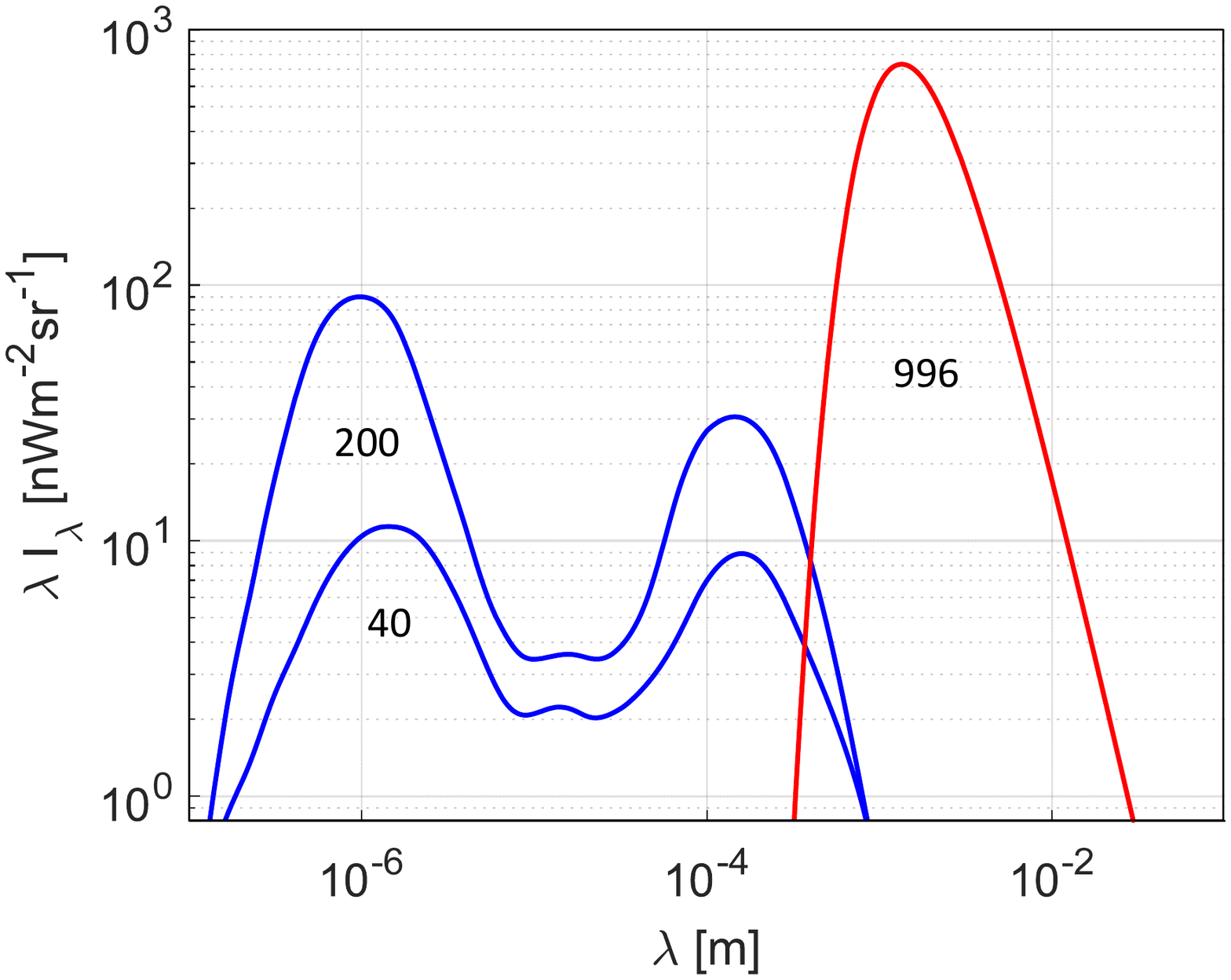}
\caption{
Spectral energy distribution of the EBL limits (blue lines) and of the CMB (red line). The EBL limits are taken from Figure~\ref{fig:1}. The numbers indicate the total intensities of the EBL and CMB.
}
\label{fig:4}
\end{figure}

The energy of light absorbed by galactic or intergalactic dust heats up the dust and produces its thermal radiation. The temperature of dust depends on the intensity of light absorbed by dust grains. Within galaxies, the light intensity is high, the galactic dust being heated up to 20-40 K and emitting a thermal radiation at infrared (IR) and far-infrared (FIR) wavelengths \citep{Schlegel1998, Draine2007}. Since the intensity of the EBL is lower than the intensity of light within galaxies, the intergalactic dust is colder and emits radiation at microwave wavelengths. At these wavelengths, the only dominant radiation is the CMB, see Figure~\ref{fig:4}.

\subsection{Absorption of EBL by intergalactic dust}

Assuming intergalactic dust to be the ideal blackbody, its temperature $T^D$ is calculated using the Stefan-Boltzmann law
\begin{equation}\label{eq14}
T^D = \left(\frac{\,\,I^D}{\pi \sigma}\right)^{\frac{1}{4}} \,,
\end{equation}
where $\sigma = 5.67 \times 10^{−8} \, \mathrm{W m}^{-2} \,\mathrm{K}^{-4}$ is the Stefan-Boltzmann constant, and $I^D$ is the total dust intensity (radiance) in $\mathrm{W m}^{-2}\,\mathrm{sr}^{-1}$. 

If we consider the thermal energy radiated by dust equal to the EBL absorbed by dust, and insert the lower and upper limits of the EBL, 40 and 200 $\mathrm{n W m}^{-2}\,\mathrm{sr}^{-1}$, obtained from observations (Figure~\ref{fig:1}) into equation (14), the temperature of the intergalactic dust ranges from 1.22 to 1.82 K. These values are much lower than the observed temperature of 2.725 K of the CMB. In order to heat up intergalactic dust to the CMB temperature, the energy flux absorbed by dust should be 996 $\mathrm{n W m}^{-2}\,\mathrm{sr}^{-1}$. This value is $5-25$ times higher than the total intensity of the EBL. Hence, if the CMB is related to the thermal radiation of intergalactic dust, the EBL forms just a fraction of the energy absorbed by dust. 

Obviously, assuming that the dust radiation is only a reprocessed EBL is not correct. A more appropriate approach should  consider the absorption of the thermal radiation of dust itself and a balance between the energy radiated and absorbed by dust and by galaxies.

\subsection{Energy balance of thermal dust radiation}

Both galaxies and intergalactic dust radiate and absorb energy. Galaxies radiate light in optical, IR and FIR spectra; intergalactic dust radiates energy in the microwave spectrum. Radiation of galaxies produces the EBL with the total intensity  $I^{\mathrm{EBL}}$, which is partly absorbed by galaxies ($I^{\mathrm{EBL}}_{AG}$) and partly by dust ($I^{\mathrm{EBL}}_{AD}$),
\begin{equation}\label{eq15}
I^{\mathrm{EBL}} = I^{\mathrm{EBL}}_{AG} + I^{\mathrm{EBL}}_{AD} \,.
\end{equation}
The same applies to dust radiation with the total intensity $I^{D}$
\begin{equation}\label{eq16}
I^{D} = I^{D}_{AG} + I^{D}_{AD} \,.
\end{equation}

\begin{figure*}
\includegraphics[angle=0,width=15cm,trim= 90 150 100 120, clip=true]{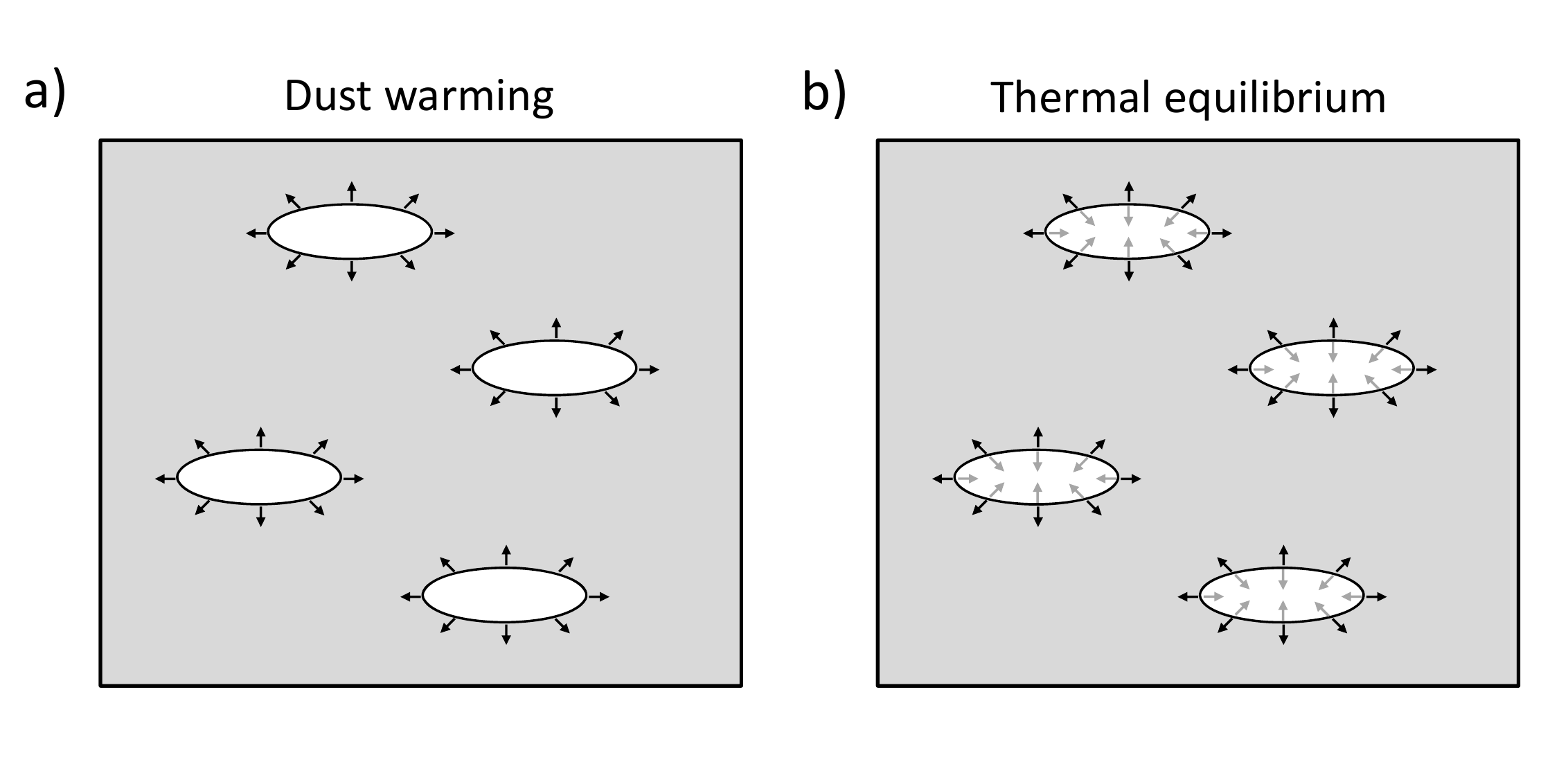}
\caption{
Scheme of the energy balance between the EBL (black arrows) and CMB (grey arrows) radiated and absorbed by galaxies (ellipses) and intergalactic dust (grey area). (a) Dust is warmed by a part of the EBL which is absorbed by dust. (b) Dust is in thermal equilibrium, when the EBL flux absorbed by intergalactic dust is compensated by the CMB flux absorbed by galaxies.
}
\label{fig:5}
\end{figure*}

If the energy radiated by dust is completely absorbed by dust (no dust radiation is absorbed by galaxies, $I^{D}_{AG}=0$) and no other sources of light are present ($I^{\mathrm{EBL}}_{AD}=0$), the dust temperature is constant.  If dust additionally absorbs some light emitted by galaxies ($I^{\mathrm{EBL}}_{AD} \ne 0$), it is being heated up and the dust temperature increases continuously with no limit (see Figure~\ref{fig:5}a). The process of heating can be terminated only if some energy emitted by dust is absorbed back by galaxies ($I^{D}_{AG} \ne 0$). In this case, dust warming continues until the intergalactic dust reaches energy equilibrium. Since the dust grains are small, the process of dust thermalization by the EBL is fast and the effect of universe expansion can be neglected. 

Under the thermal equilibrium of dust, the energy interchanged between galaxies and dust is mutually compensated
\begin{equation}\label{eq17}
I^{\mathrm{EBL}}_{AD} = I^{D}_{AG} \,,
\end{equation}
and the total energy of dust is conserved (see Figure~\ref{fig:5}b). 

Since the proportion between the energy absorbed by intergalactic dust and by galaxies is controlled by the opacity ratio
\begin{equation}\label{eq18}
I^{\mathrm{EBL}}_{AD} = R_\kappa I^{\mathrm{EBL}}_{AG}\, , \,  I^{D}_{AD} = R_\kappa I^{D}_{AG} \,,
\end{equation}
we can rewrite equations (15) and (16) to read
\begin{equation}\label{eq19}
I^{\mathrm{EBL}} = \frac{1+R_\kappa}{R_\kappa} I^{\mathrm{EBL}}_{AD} \,, \,
I^{D} = \left(1+R_\kappa\right) I^{D}_{AG} \,\, ,
\end{equation}
and the relation between the intensity of dust radiation and the EBL is finally expressed using equation (17) as
\begin{equation}\label{eq20}
I^{D} = R_\kappa  I^{\mathrm{EBL}} \,,
\end{equation}
where $R_\kappa$ is defined in equation (12) and estimated in Table~\ref{Table:1}. 

Equation (20) is invariant to the cosmological model considered and its validity can be verified by observations. The EBL intensity estimated using current measurements ranges from 40 to $200 \, \mathrm{n W m}^{-2}\,\mathrm{sr}^{-1}$ (see Figure~\ref{fig:4}) with an optimum value of about $80 \, \mathrm{n W m}^{-2}\,\mathrm{sr}^{-1}$. The optimum dust temperature predicted from equation (20) when inserting $80 \, \mathrm{n W m}^{-2}\,\mathrm{sr}^{-1}$ for the $I^{\mathrm{EBL}}$ and 13.4 for the $R_\kappa$ is 
\begin{equation}\label{eq21}
T^{D}_{\mathrm{theor}} = 2.776 \, \mathrm{K} \,,
\end{equation}
which is effectively the CMB temperature. The difference between the predicted temperature of dust and the observed CMB temperature is about 1.9\% being caused by inaccuracies in the estimates of the EBL intensity and of the opacity ratio.

If we substitute the predicted dust intensity $I^{D}$ corresponding to temperature 2.776 K by the CMB intensity $I^{\mathrm{CMB}}$ corresponding to temperature 2.725 K in equation (20), we can calculate the opacity ratio $R_\kappa$ defined in equation (12) in the following alternative way:
\begin{equation}\label{eq22}
R_\kappa = \frac{I^{\mathrm{CMB}}}{I^{\mathrm{EBL}}} \,.
\end{equation}
This ratio lies in the range of $5-25$ with the optimum value of 12.5 which is quite close to the value of 13.4 obtained from measurements of the galactic and intergalactic opacity using equation (12), see Table~\ref{Table:1}.

\subsection{Redshift dependence of the dust temperature and dust radiation}

The thermal radiation of the intergalactic dust must depend on redshift similarly as any radiation in the expanding universe. The redshift dependence of the intensity of dust radiation is derived from equation (20). Since the opacity ratio $R_\kappa$ does not depend on redshift and the redshift dependence of $I^{\mathrm{EBL}}$  is described by equation (5), we get
\begin{equation}\label{eq23}
I^{D}\left(z\right) = \left(1+z\right)^4 I_0^{D}  \,,
\end{equation}
where $I_0^{D}$ is the intensity of dust radiation at redshift $z = 0$.

Inserting equation (23) into equation (14) the dust temperature at redshift $z$ comes out
\begin{equation}\label{eq24}
T^{D}\left(z\right) = \left(1+z\right) T_0^{D} \,,
\end{equation}
where $T_0^{D}$ is the temperature of dust at $z = 0$. Hence, the dust temperature linearly increases with redshift $z$. Similarly as the Tolman relation, equation (24) is invariant to the cosmological model applied, being based only on the assumptions of conservation of the galaxy number density, dust density and constant galaxy luminosity in the comoving volume. 

\section{Spectral and total intensity of dust radiation}

\subsection{Spectral intensity of dust radiation}

If we assume dust to be the blackbody, its thermal radiation (i.e., energy emitted per unit projected area into a unit solid angle in the frequency interval $\nu$ to $\nu + d\nu$) is described by the Planck's law
\begin{equation}\label{eq25}
I_\nu\left(\nu,T^D\right) = \frac{2h\nu^3}{c^2} \frac{1}{e^{h\nu/k_BT^D}-1} \,\,\,(\mathrm{in \, W m}^{-2}\,\mathrm{sr}^{-1} \mathrm{Hz}^{-1}) \,,
\end{equation}
where $\nu$ is the frequency, $T^D$ is the dust temperature, $h$ is the Planck constant, $c$ is the speed of light, and $k_B$ is the Boltzmann constant. The dust temperature is uniform for a given time instant, but increases with redshift $z$, see equation (24). Since the received wavelengths also increase with redshift $z$, we arrive at
\begin{equation}\label{eq26}
\frac{h\nu}{k_B T^D} =  \frac{h\nu_0 \left(1+z\right)}{k_B T_0^D \left(1+z\right)} = \frac{h\nu_0}{k_B T_0^D} \,.
\end{equation}
Hence, the temperature increase with $z$ exactly eliminates the frequency redshift of the thermal radiation in equation (25). Consequently, the radiation of dust observed at all distances looks apparently as the radiation of the blackbody with a single temperature.

\subsection{Total intensity of dust radiation}
Assuming the temperature of dust particles at redshift $z$ to be 
\begin{equation}\label{eq27}
T^{D}\left(z\right) = \left(1+z\right) T_0^{\mathrm{CMB}} \,,
\end{equation}
we can calculate the total intensity of thermal dust radiation. 

The total (bolometric) intensity $I^D$ of the dust radiation (in $\mathrm{W m}^{-2}\, \mathrm{sr}^{-1}$) is expressed as an integral over redshift $z$ 
\begin{equation}\label{eq28}
I^D = \frac{1}{4 \pi} \int_0^{\infty} j^D\left(z\right)  
\,e^{-\tau^D \left(z\right)}\, \frac{c}{H_0} \frac{dz}{E\left(z\right)} \,,
\end{equation}
where $j^D\left(z\right)$ is the luminosity density of dust radiation, and $\tau^D \left(z\right)$ is the effective optical depth of the Universe at the CMB wavelengths produced by intergalactic dust. The term $ {1/\left(1+z\right)^2}$ expressing the reduction of the received energy caused by redshift $z$ and present, for example, in a similar formula for the EBL \citep[his equation 1]{Vavrycuk2017a} is missing in equation (28), because the energy reduction is eliminated by the redshift dependence of dust temperature. Since the temperature increases linearly with $z$, the dust luminosity density $j^D\left(z\right)$ reads
\begin{equation}\label{eq29}
j^{D}\left(z\right) = \left(1+z\right)^4 j_0^D \,,
\end{equation}
where $j_0^D$ is the dust luminosity density in the local Universe ($z = 0$). Consequently,
\begin{equation}\label{eq30}
I^D = \frac{j_0^D}{4 \pi} \int_0^{\infty} \left(1+z\right)^4  
\,e^{-\tau^D \left(z\right)}\, \frac{c}{H_0} \frac{dz}{E\left(z\right)} \,.
\end{equation}
The effective optical depth $\tau^D \left(z\right)$ reads
\begin{equation}\label{eq31}
\tau^D\left(z\right) = \frac{c}{H_0} \int_0^{z} \lambda_0^D \left(1+z'\right)^4 \,\, \frac{dz'}{E\left(z'\right)} \,,
\end{equation}
where $\lambda_0^D$ is the mean intergalactic absorption coefficient of dust radiation along the ray path. The term describing the absorption of the CMB by galaxies is missing in equation (31) because it is exactly compensated by the EBL radiated by galaxies and absorbed by intergalactic dust, see equation (17).

Taking into account the following identity
\begin{equation}\label{eq32}
\int_0^{\infty} f\left(z\right)\, \mathrm{exp}\left(-\int_0^{z} f\left(z'\right) dz' \right) dz = 1 \,,
\end{equation}
assuming $f\left(z\right) \rightarrow \infty$ for $z \rightarrow \infty$, the intensity of dust radiation comes out as
\begin{equation}\label{eq33}
I^D = \frac{1}{4 \pi} \frac{j_0^D}{\lambda_0^D} \,.
\end{equation}
Since the luminosity density of dust radiation $j_0^D$ reads
\begin{equation}\label{eq34}
j_0^D = n_0^D E^D = 4 n_0^D \sigma^D L^D = 4 \pi n_0^D \sigma^D I^{\mathrm{CMB}} \,,
\end{equation}
where $n_0^D$ is the number density of dust particles at $z = 0$, $E^D$ is the total luminosity of one dust particle (in W), $\sigma^D$ is the mean cross-section of dust particles, $L^D$ is the energy flux radiated per unit surface of dust particles (in Wm$^{-2}$), and $I^\mathrm{CMB}$ is the intensity radiated by a blackbody with the CMB temperature (in $\mathrm{W m}^{-2}\, \mathrm{sr}^{-1}$). Since 
\begin{equation}\label{eq35}
\lambda_0^D = n_0^D \sigma_D \,,
\end{equation}
equations (33) and (34) yield
\begin{equation}\label{eq36}
I^D = I^{\mathrm{CMB}} \,.
\end{equation}
Equation (36) is valid independently of the cosmological model considered and states that the energy flux received by the unit area of the intergalactic space is equal to the energy flux emitted by the unit area of intergalactic dust particles. This statement is basically a formulation of the Olbers' paradox \citep[his equation 9]{Vavrycuk2016} applied to dust particles instead of to stars. Since the sky is fully covered by dust particles and distant background particles are obscured by foreground particles, the energy fluxes emitted and received by dust are equal. This is valid irrespective of the actual dust density in the local Universe.

\section{Saturation redshift of CMB}

\begin{figure*}
\includegraphics[angle=0,width=13cm,trim=100 30 120 40, clip=true]{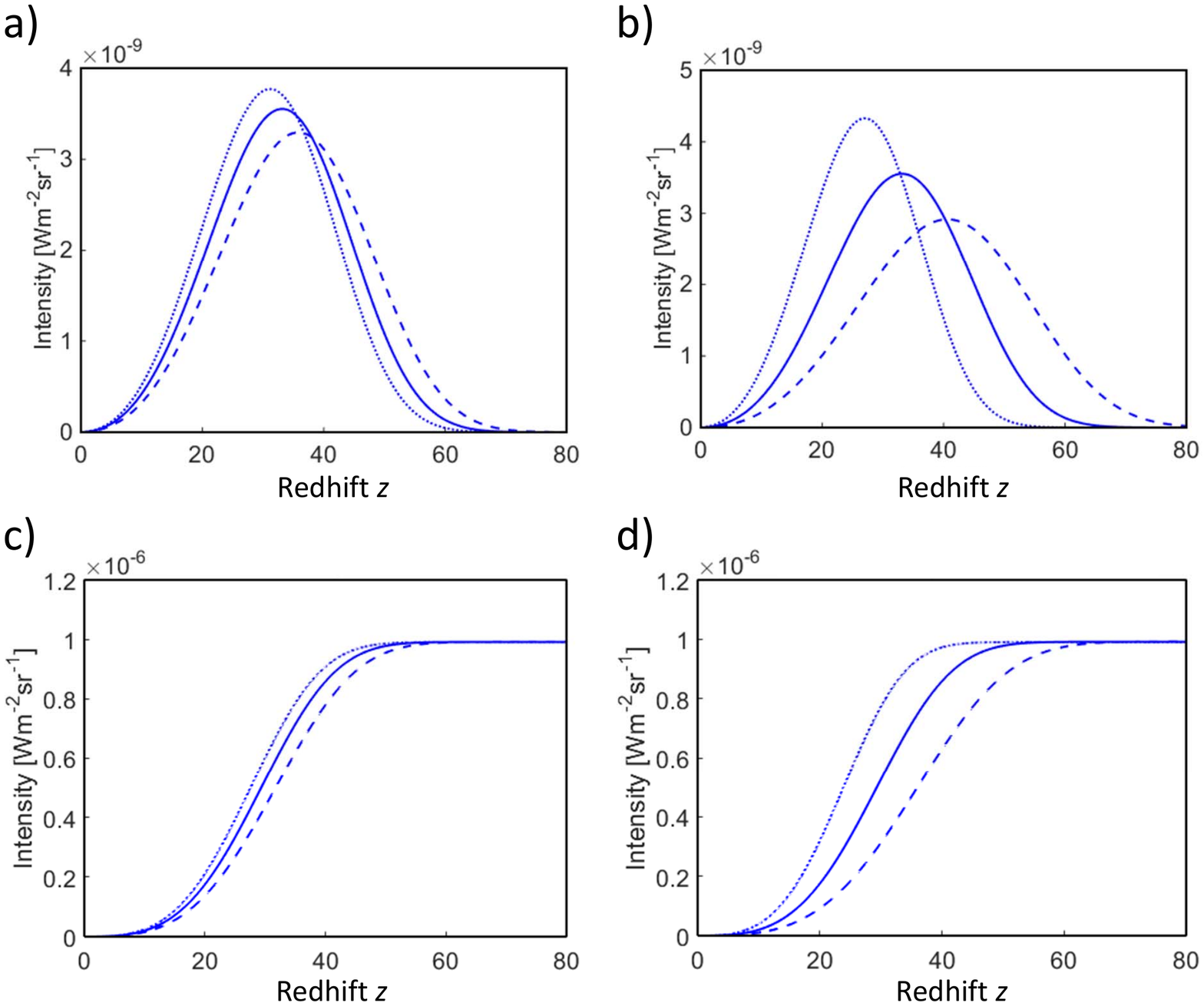}
\caption{
The intensity of the CMB radiated by dust at redshift $z$ (a-b), and the cumulative intensity of the CMB emitted by dust up to redshift $z$ (c-d). 
(a) and (c) - intergalactic attenuation $A_V$ is $1.5 \times 10^{-2}$ (blue dashed line), $2.0 \times 10^{-2}$ (blue solid line) and $2.5 \times 10^{-2}\, \mathrm{mag}\, h \, \mathrm{Gpc}^{-1}$ (blue dotted line). The ratio between the CMB and visual attenuations is $1.0 \times 10^{-4}$. 
(b) and (d) - the ratio between the CMB and visual attenuations is $0.5 \times 10^{-4}$ (blue dashed line), $1.0 \times 10^{-4}$ (blue solid line) and $2.0 \times 10^{-4}$ (blue dotted line).  The intergalactic attenuation $A_V$ is $2.0 \times 10^{-2}\, \mathrm{mag}\, h \, \mathrm{Gpc}^{-1}$. Cosmological parameters:  $H_0 = 67.7 \,\mathrm{km}\,\mathrm{s}^{-1} \mathrm{Mpc}^{-1}$, $\Omega_m = 0.3$, and $\Omega_\Lambda = 0.7$. 
}
\label{fig:6}
\end{figure*}

The total intensity of the CMB is calculated by summing the intensity over all redshifts $z$, see equation (30). Since the intensity is attenuated due to the exponential term with the optical depth in equation (30), the contribution of the dust radiation to the energy flux decreases with redshift $z$. Since most of the CMB energy is absorbed by intergalactic dust but not by galaxies, the optical depth depends basically on the attenuation of intergalactic dust at the CMB wavelengths $\lambda^{\mathrm{CMB}}$, which can be expressed as
\begin{equation}\label{eq37}
\lambda^{\mathrm{CMB}} = k^{\mathrm{CMB}} \lambda_V \,,
\end{equation}
where $k^{\mathrm{CMB}}$ is the ratio between the attenuation of intergalactic dust at the CMB and visual wavelengths. The lower the CMB attenuation, the slower the decrease of dust radiation with redshift. Consequently, we can define the so-called saturation redshift $z^*$ as the redshift for which the CMB intensity reaches 98\% of its total value:
\begin{equation}\label{eq38}
I^D\left(z^*\right) = \frac{j_0^D}{4 \pi} \int_0^{z^*} \left(1+z\right)^4  
\,e^{-\tau^D \left(z\right)}\, \frac{c}{H_0} \frac{dz}{E\left(z\right)} = 0.98 \, I^D \,.
\end{equation}
Assuming that the expanding history of the Universe is correctly described by equation (38) for redshifts up to 50-60 and inserting $0.02 \, \mathrm{mag} \, h \, \mathrm{Gpc}^{-1}$ for the visual intergalactic opacity and $1 \times 10^{-4}$ for the ratio between the CMB and visual extinctions (see Figure~\ref{fig:2}), we get the CMB to be saturated at redshifts of about $z^* = 55$ (see Figure~\ref{fig:6}). If the ratio is lower by one order, the saturation redshift is about 100. A rather high value of the CMB saturation redshift indicates that the observed CMB intensity is a result of dust radiation summed over vast distances of the Universe. As a consequence, the CMB intensity must be quite stable with only very small variations with direction in the sky.

\section{CMB temperature anisotropies}

\begin{figure*}
\includegraphics[angle=0,width=16.0cm,trim = 60 80 60 70, clip=true]{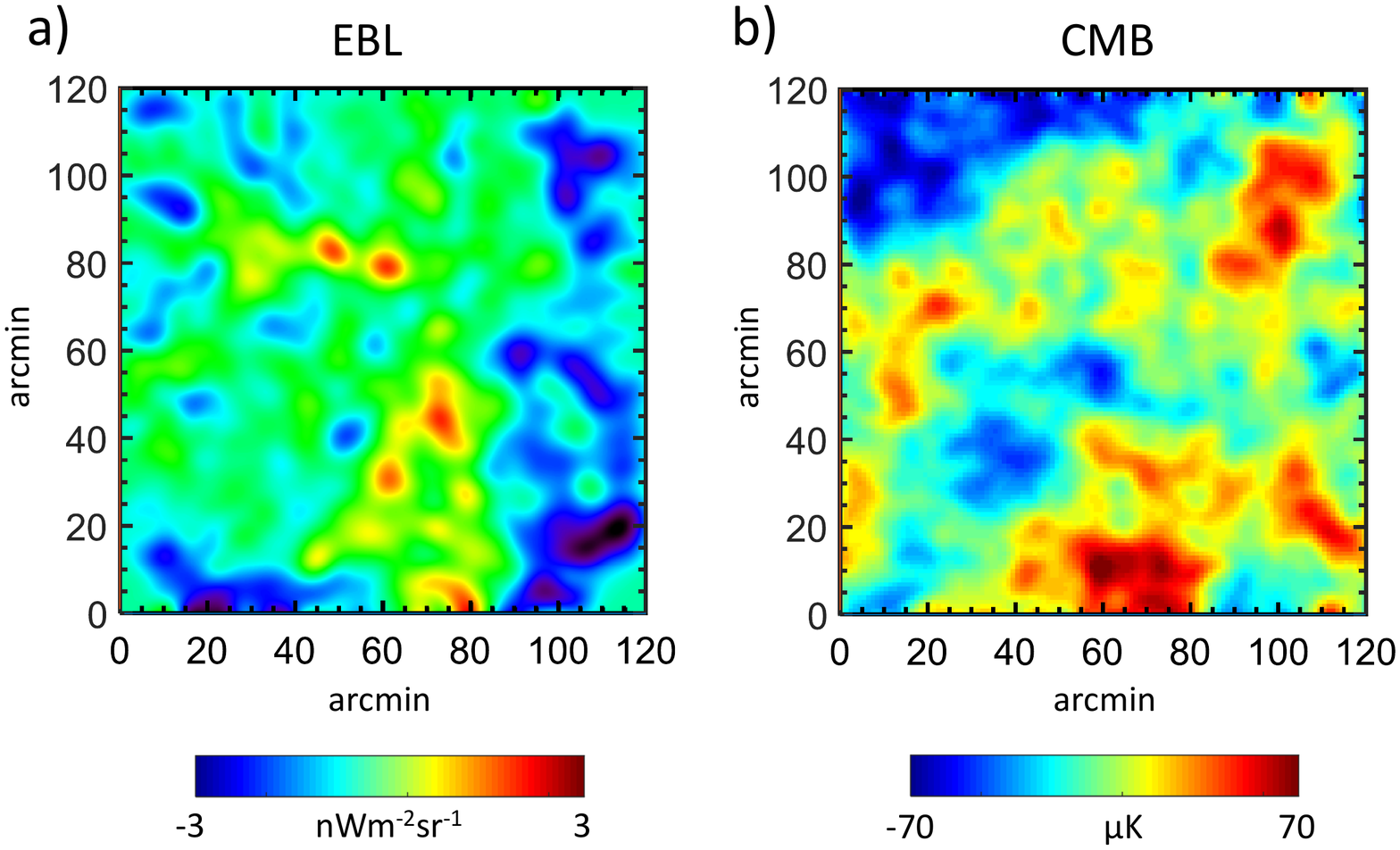}
\caption{
Examples of the angular fluctuations of (a) the specific intensity of the EBL, and (b) the CMB temperature. The EBL plot is taken from \citet[their Figure S8]{Zemcov2014} and represents a differenced 1.1 $\mu$m image of two fields (NEP - ELAIS-N1) smoothed with the Gaussian function with FWHM of 7.2' to highlight a large-scale structure. The differenced image is used by \citet{Zemcov2014} to reduce the effect of flat-fielding errors. The CMB plot was produced using tools of the NASA/IPAC Infrared Science Archive and represents a smoothed image of the Planck public intensity data (Release 2) at 353 GHz for the ELAIS-N1 field at 16h11m.5, 54d37m.9. The image is rescaled to temperature. Since the EBL plot does not show an actual field but a differenced field, the patterns of the EBL and CMB plots indicate a rough similarity but they cannot mutually correlate.
}
\label{fig:7}
\end{figure*}

So far, we have assumed the EBL to be perfectly isotropic with no directional variation, being a function of redshift only. Obviously, this assumption is not correct because of clustering of galaxies and presence of voids in the Universe \citep{Bahcall1999, Hoyle2004, Jones2004, vonBenda_Beckmann2008, Szapudi2015}. Consequently, the EBL displays fluctuations (Figure~\ref{fig:7}a) manifested as small-scale EBL anisotropies reported mostly at IR wavelengths \citep{Kashlinsky2002, Cooray2004, Matsumoto2005, Kashlinsky2007, Matsumoto2011, Cooray2012, Pyo2012, Zemcov2014}.  Since the EBL forms about 7\% of the total energy absorbed by dust and reradiated as the CMB, the EBL fluctuations should affect the intensity of the CMB (Figure~\ref{fig:7}b). The remaining 93\% of the total energy absorbed by dust is quite stable because it comes from the CMB itself which is averaged over large distances.  

Let the luminosity density of dust radiation $j^D$ display small-scale variations with distance $\Delta\left(r\right)$ reflecting the EBL fluctuations in the Universe. Transforming distance to redshift and taking into account the redshift dependence of $j^D$, we get 
\begin{equation}\label{eq39}
j^D\left(z\right) = \left(1+\Delta \left(z\right)\right) \left(1+z\right)^4j_0^D \,,
\end{equation}
where $\Delta(z)$ is much smaller than 1 and has a zero mean value. Inserting equation (39) into equation (28), the variation of the total intensity of the dust radiation $\Delta I^D$  (in $\mathrm{W m}^{-2}\,\mathrm{sr}^{-1}$) reads
\begin{equation}\label{eq40}
\Delta I^D = \frac{j_0^D}{4 \pi} \int_0^{\infty} \Delta\left(z\right) \left(1+z\right)^4  
\,e^{-\tau^D \left(z\right)}\, \frac{c}{H_0} \frac{dz}{E\left(z\right)}  \,,
\end{equation}
where the optical depth $\tau^D\left(z\right)$ is defined in equation (31). The variation of the CMB temperature corresponding to $\Delta I^D$ is obtained using equation (14).

\begin{figure}
\includegraphics[angle=0,width=8.5cm,trim= 200 140 150 150, clip=true]{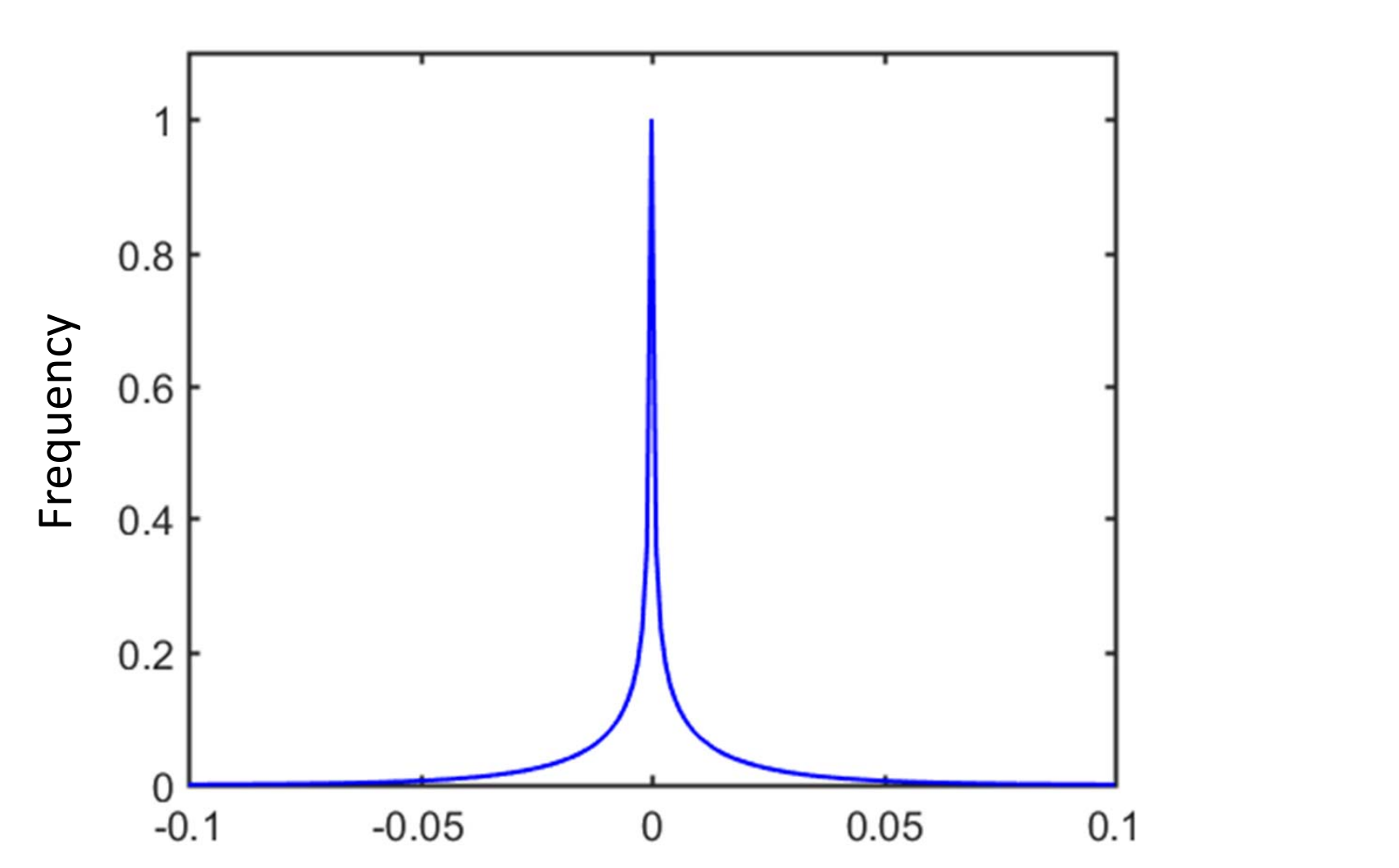}
\caption{
The probability density function of the modelled luminosity density fluctuations of dust radiation.
}
\label{fig:8}
\end{figure}

\begin{figure*}
\includegraphics[angle=0,width=15cm,trim=90 210 90 120, clip=true]{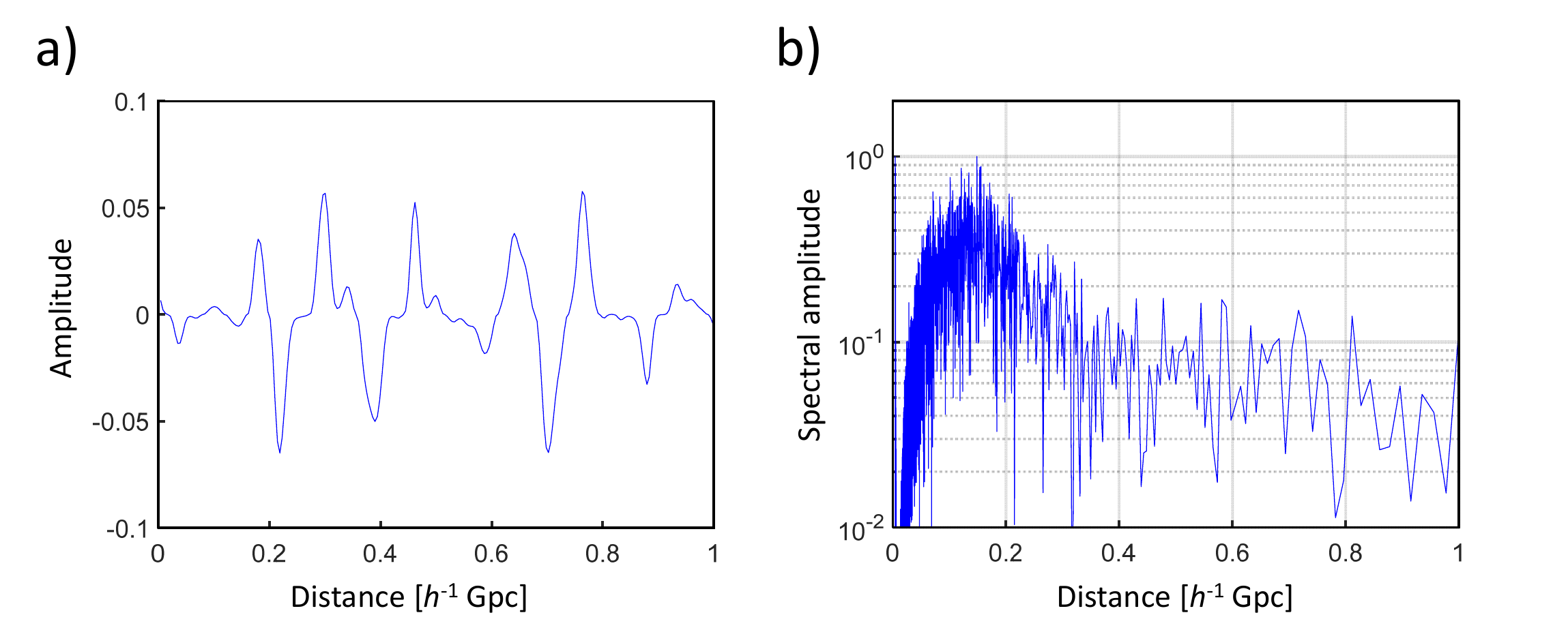}
\caption{
The luminosity density fluctuations of dust radiation (a) and its spectrum (b) as a function of distance. 
}
\label{fig:9}
\end{figure*}

The sensitivity of the intensity of the dust radiation and the dust temperature to the EBL fluctuations can be tested numerically. Figures~\ref{fig:8} and~\ref{fig:9} show synthetically generated fluctuations of the luminosity density $\Delta\left(r\right)$. The fluctuations were generated by bandpass filtering of white noise. To mimic real observations, we kept predominantly fluctuations of size between 20-100 Mpc, corresponding to typical cluster, supercluster and void dimensions \citep{Bahcall1999, Hoyle2004, vonBenda_Beckmann2008}. The other sizes of fluctuations were suppressed. The probability density function of $\Delta\left(r\right)$ is very narrow with a standard deviation of 0.02 (Figure~\ref{fig:8}). The noise level of 2\% expresses the fact that the variations of the EBL should contribute to the dust radiation by less than 10\% only. 

Considering the luminosity density fluctuations shown in Figure~\ref{fig:9} and the intergalactic opacity at the CMB wavelengths $1 \times 10^{-4}$ lower than that at visual wavelengths (see Fig. 2) in modelling of the intensity variation  $\Delta I^D$ using equation (40), we find that $\Delta I^D$  attains values up to $\pm 0.25\, \mathrm{n W m}^{-2}\,\mathrm{sr}^{-1}$ and the standard deviation of dust temperature is $\pm 60\, \mu \mathrm{K}$. The maximum variation of the dust temperature is up to $\pm 170\, \mu \mathrm{K}$. The standard deviations and the maximum limits were obtained from 1000 noise realizations. Compared to observations, the retrieved variations are reasonable. Taking into account very rough estimates of input parameters, the predicted temperature variation fits well the observed small-scale anisotropies of the CMB attaining values up to $\pm 70\, \mu \mathrm{K}$ \citep{Bennett2003, Hinshaw2009, Bennett2013, Ade2014a, Ade2014b}. More accurate predictions are conditioned by a detailed mapping of EBL fluctuations by planned cosmological missions such as Euclid, LSST or WFIRST  \citep{vanDaalen2011,Masters2015}. For example, the NASA explorer SPHEREx will be able to conduct a three-dimensional intensity mapping of spectral lines such as H$\alpha$ at $z \sim 2$ and Ly$\alpha$ at $z>6$ over large areas in the sky \citep{Cooray2016, Fonseca2017, Gong2017}.

The relation between the presence of voids, clusters and the CMB anisotropies predicted in this paper has been recently supported by observations of several authors. The studies were initiated by detecting an extreme cold spot (CS) of $-70 \, \mu$K in the WMAP images \citep{Vielva2004, Cruz2005} which violated a condition of Gaussianity of the CMB required in the Big Bang theory. Later, the origin of the CS was attributed to the presence of a large void detected by the NRAO VLA Sky Survey \citep{Rudnick2007}. The presence of the supervoid was later confirmed by \citet{Gurzadyan2014} using the Kolmogorov map of Planck's 100 GHz data and by \citet{Szapudi2015} using the WISE-2MASS infrared galaxy catalogue matched with Pan-STARRS1 galaxies. The radius of the supervoid was estimated by \citet{Szapudi2015} to be $R_{\mathrm{void}} = 220 \pm 50 \, h^{-1} \, \mathrm{Mpc}$. 

A physical relation between large-scale structures and the CMB anisotropies has been confirmed also for other spots. For example, a large low-density anomaly in the projected WISE-2MASS galaxy map called the Draco supervoid was aligned with a CMB decline by \citet{Finelli2016}. The imprint of superstructures on the CMB was also statistically evidenced by stacking CMB temperatures around the positions of voids from the SDSS DR7 spectroscopic galaxy catalogue \citep{Cai2014}. In addition, \citet{Kovacs2017} probed the correlation between small temperature anomalies and density perturbations using the data of the Dark Energy Survey (DES). They identified 52 large voids and 102 superclusters at redshifts $0.2 < z < 0.65$ and performed a stacking measurement of the CMB temperature field based on the DES data. They detected a cumulative cold imprint of voids with $\Delta T = −5.0 \pm 3.7 \, \mu K$ and a hot imprint of superclusters $\Delta T = 5.1 \pm 3.2 \, \mu K$.

\begin{figure}
\includegraphics[angle=0,width=10cm,trim=170 60 120 10, clip=true]{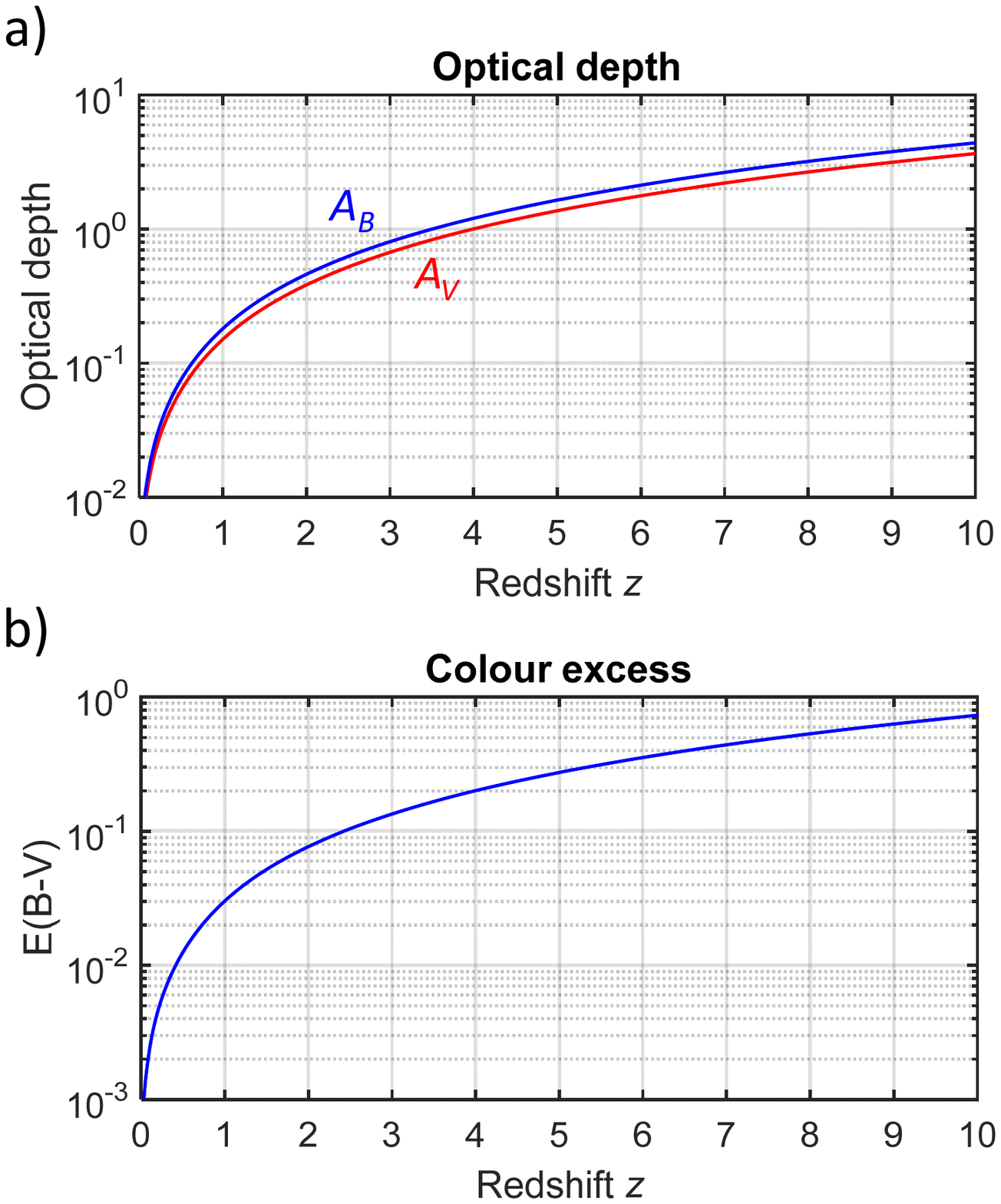}
\caption{
Optical depth and colour excess of intergalactic space as a function of redshift. The extinction coefficient $R_V$ is assumed to be 5. $A_V$ - extinction at the visual band, $A_B$ - extinction at the B band. 
}
\label{fig:10}
\end{figure}

\section{CMB polarization}

Small-scale CMB anisotropies are expressed not only in temperature but also in polarization being mutually correlated. In the dust theory, the CMB polarization can be explained by interaction of dust with a cosmic magnetic field produced by large-scale structures in the Universe. Magnetic fields are present in all types of galaxies, clusters and superclusters. The Milky Way has a typical interstellar magnetic field strength of 2 $\mu$G in a regular ordered component on kpc scales \citep{Kulsrud1999}. Other spiral galaxies have magnetic field strengths of 5 to 10 $\mu$G, with field strengths up to 50 $\mu$G in starburst galaxy nuclei \citep{Beck1996}. The magnetic fields in galaxy clusters and in the intergalactic medium (IGM) have strength of $10^{-5}-10^{-7}$ G \citep{Widrow2002, Vallee2004, Giovannini2004}. They are present even in voids with a minimum strength of about $10^{-15}$ G on Mpc scales \citep{Beck2013}. The intercluster magnetic fields have been measured using synchrotron relic and halo radio sources within clusters, inverse Compton X-ray emission from clusters, surveys of Faraday rotation measures of polarized radio sources both within and behind clusters, and studies of cluster cold fronts in X-ray images \citep{Carilli2002}. The measurements suggest substantially magnetized atmospheres of most clusters. Models of the magnetic field of the IGM typically involve an ejection of the fields from normal or active galaxies \citep{Heckman2001}. \citet{Kronberg1999} considered this mechanism and showed that a population of dwarf starburst galaxies at $z \geq 6$ could magnetize almost 50\% of the universe.

The magnetic fields are easily traced via polarization of radiation resulting from extinction or/and emission by aligned dust grains in the interstellar or intergalactic medium \citep{Lazarian2007, Andersson2015}. The grain alignment by magnetic fields proved to be an efficient and rapid process which causes a linear polarization of starlight when passing through the dust and a polarized thermal emission of dust \citep{Lazarian2007}. Hence, optical, IR and FIR polarimetry can reveal the presence and the detailed structure of magnetic fields in our Galaxy \citep{Ade2015, Aghanim2016b} as well as in large-scale formations in the Universe \citep{Feretti2012}. The polarized light can be used not only for tracing magnetic fields but also for detecting dust. For example, the observation of submillimetre polarization proves a needle origin for the cold dust emission and presence of metallic needles in the ejecta from core-collapse supernovae \citep{Dwek2004, Dunne2009}.

Tracing magnetic fields in our Galaxy is particularly important for the analysis of the CMB because the polarized galactic dust forms a foreground which should be eliminated from the CMB polarization maps \citep{Lazarian2002, Gold2011, Ichiki2014, Ade2015, Aghanim2016b}. Some authors also point to a possible interaction of the CMB with intergalactic magnetic fields \citep{Ohno2003} and admit that these fields may modify the pattern of the CMB anisotropies and, eventually, induce additional anisotropies in the polarization \citep{Giovannini2004}. However, since the CMB is believed to be a relic radiation originating in the Big Bang, the possibility that the CMB is actually a dust radiation with polarization tracing the large-scale magnetic fields has not been investigated or proposed. 

Assuming that the CMB is produced by thermal radiation of intergalactic dust, the small-scale polarization anisotropies of the CMB are readily explained by the polarized thermal radiation of needle-shaped conducting dust grains present in the IGM and aligned by cosmic magnetic fields produced by large-scale structures in the Universe. The phenomenon is fully analogous to the polarized interstellar dust emission in the Milky Way, which is observed at shorter wavelengths because the temperature of the interstellar dust is higher than that of the intergalactic dust. Since both the temperature and polarization anisotropies of the CMB are caused by clusters and voids in the Universe, they are spatially correlated.

\section{Dust in the high-redshift Universe}

The presence of a significant amount of dust is unexpected at high redshifts in the Big Bang theory but reported in recent years by many authors. For example, a submillimitre radiation coming from warm dust in the quasar host galaxy is detected for a large number of high-redshift quasars ($z > 5-6$) observed by the IRAM 30-m telescope or SCUBA \citep{Priddey2003, Fan2006, Priddey2008} or mm and radio radiation of quasars observed with the Max Planck Millimetre Bolometer Array (MAMBO) at 250 GHz \citep{Wang2008}. Similarly, the existence of mature galaxies in the early Universe indicates that this epoch was probably not as dark and young as so far assumed. Based on observations of the Atacama Large Millimetre Array (ALMA), \citet{Watson2015} investigated a galaxy at $z > 7$ highly evolved with a large stellar mass and heavily enriched in dust. Similarly, \citet{Laporte2017} analysed a galaxy at a photometric redshift of $z \sim 8$ with a stellar mass of $\sim 2 \times 10^{9} \, M_{\sun}$, a SFR of $\sim 20 M_{\sun}/\mathrm{yr}$ and a dust mass of $\sim 6 \times 10^{6} M_{\sun}$. Also a significant increase in the number of galaxies for $8.5 < z < 12$ reported by \citet{Ellis2013} and the presence of a remarkably bright galaxy at $z \sim 11$ found by \citet{Oesch2016} questions the assumption of the age and darkness of the high-redshift Universe. 

Although numerous recent observations confirm significant reddening of galaxies and quasars caused by the presence of dust at high redshifts, it is unclear which portion of the reddening is produced by local dust in a galaxy and by intergalactic dust along the line of sight. \citet{Xie2015, Xie2016} tried to distinguish between both sources of extinction by studying spectra of $\sim 90.000$ quasars from the SDSS DR7 quasar catalogue \citep{Schneider2010}. They calculated composite spectra in the redshift intervals $0.71 < z < 1.19$ and $1.80 < z < 3.15$ in four bolometric luminosity bins and revealed that quasars at higher redshifts have systematically redder UV continuum slopes indicating intergalactic extinction $A_V$ of about $2 \times 10^{-5}\, h\, \mathrm{Mpc}^{-1}$, see Figure~\ref{fig:10} for its redshift dependence.  

The dust content in the IGM can also be probed by studying absorption lines in spectra of high-redshift quasars caused by intervening intergalactic gaseous clouds. The massive clouds reach neutral hydrogen column densities $N_\mathrm{HI} > 10^{19} \, \mathrm{cm}^{-2}$ (Lyman-limit systems, LLS) or even $N_\mathrm{HI} > 2 \times 10^{20} \, \mathrm{cm}^{-2}$ (damped Lyman systems, DLA) and they are self-shielded against ionizing radiation from outside \citep{Wolfe2005, Meiksin2009}. They have higher metallicities than any other class of Lyman absorbers ([M/H] $\sim$ -1.1 dex; \citet{Pettini1994} and they are expected to contain dust. The dust content is usually estimated from the abundance ratio [Cr/Zn] assuming that this stable ratio is changed in dusty environment because Cr is depleted on dust grains but Zn is undepleted. For example, \citet{Pettini1994} analysed the [Cr/Zn] ratio of 17 DLAs at $z_\mathrm{abs} \sim 2$ and reported a typical dust-to-gas ratio of 1/10 of the value in the interstellar medium (ISM) in our Galaxy. Another analysis of 18 DLAs at $0.7 < z_\mathrm{abs} < 3.4$ performed by \citet{Pettini1997} yielded a dust-to-gas ratio of about 1/30 of the Galaxy value. Other dust indicators such as depletion of Fe and Si relative to S and Zn were used by \citet{Petitjean2002} who studied a dust pattern for a DLA at $z_\mathrm{abs} = 1.97$ and provided evidence for dust grains with an abundance similar to that in the cold gas in the Galaxy.

Similarly as for the interstellar medium, dust also well correlates with the molecular hydrogen $\mathrm{H}_2$ in clouds \citep{Levshakov2002, Petitjean2008}, because  $\mathrm{H}_2$ is formed on the surfaces of dust grains \citep{Wolfe2005}. Since the molecular hydrogen fraction sharply increases above a H I column density of $5 \times 10^{20} \, \mathrm{cm}^{-2}$ \citep{Noterdaeme2017}, DLAs can be expected to form reservoirs of dust. For example, \citet{Noterdaeme2017} discovered a molecular cloud in the early Universe at $z_\mathrm{abs} = 2.52$ with a supersolar metallicity and an overall molecular hydrogen fraction of about 50\%, which contained also carbon monoxide molecules. The authors suggest the presence of small dust grains to explain the observed atomic and molecular abundances.

\section{Corrections of luminosity and stellar mass density for intergalactic dust}

The presence of dust in the high-redshift Universe has consequences for determining the star formation rate (SFR) and the global stellar mass history (SMH). So far observations of the SFR and SMH are based on measurements of the luminosity density evolution at UV and NIR wavelengths assuming a transparent universe. The most complete measurements of the luminosity density evolution are for the UV luminosity based on the Lyman break galaxy selections covering redshifts up to 10-12 \citep{Bouwens2011, Bouwens2015, Oesch2014}. The measurements of the global stellar mass density require surveys covering a large fraction of the sky such as the SDSS and 2dFGRS. The local stellar mass function was determined, for example, by \citet{Cole2001} using the 2dFGRS redshifts and the NIR photometry from the 2MASS, and by \citet{Bell2003} from the SDSS and 2MASS data. The observed luminosity density averaged over different types of galaxies steeply increases with redshift as $(1 + z)^4$ for $z$ less than 1 \citep{Franceschini2008, Hopkins2004}. The luminosity density culminates at redshifts of about 2-3 and then decreases. The stellar mass density displays no significant evolution at redshifts $z < 1$ \citep{Brinchmann2000, Cohen2002}. However, a strong evolution of the stellar mass density is found at higher redshifts, $1 < z < 4$, characterized by a monotonous decline \citep{Hopkins2006, Marchesini2009}. This decline continues even for redshifts $z > 4$ \citep{Gonzalez2011, Lee2012}.

Obviously, if a non-zero intergalactic opacity is considered, the observations of the SFR and SMH must be corrected. The apparent stellar mass density $\rho \left(z\right)$ determined under the assumption of a transparent universe and the true stellar mass density $\rho_{\mathrm{true}} \left(z\right)$ determined for a dusty universe are related as 
\begin{equation}\label{eq41}
\rho_{\mathrm{true}} \left(z\right) = \rho \left(z\right) e^{\tau\left(z\right)} \,,
\end{equation}
where $\tau\left(z\right)$ is the optical depth of intergalactic space \citep[his equation 19]{Vavrycuk2017a}
\begin{equation}\label{eq42}
\tau\left(z\right) = \frac{c}{H_0} \int_0^{z} \lambda_0 \left(1+z'\right)^2 \,\, \frac{dz'}{E\left(z'\right)} \,,
\end{equation}
and $\lambda_0$ is the UV intergalactic attenuation at zero redshift. 

Analogously, the apparent UV luminosity density $j$ at redshift $z$ is corrected for dust attenuation by multiplying it with the exponential factor $e^{\tau\left(z\right)}$ 
\begin{equation}\label{eq43}
j_{\mathrm{true}} \left(z\right) = j\left(z\right)\,\left(1+z\right)^{-3}e^{\tau\left(z\right)} \,.
\end{equation}

The additional term $(1+z)^{-3}$ in equation (43) originates in the transformation from the comoving to proper volumes (for a detailed discussion, see Appendix A).

\begin{figure*}
\includegraphics[angle=0,width=16.5cm,trim=60 140 80 140, clip=true]{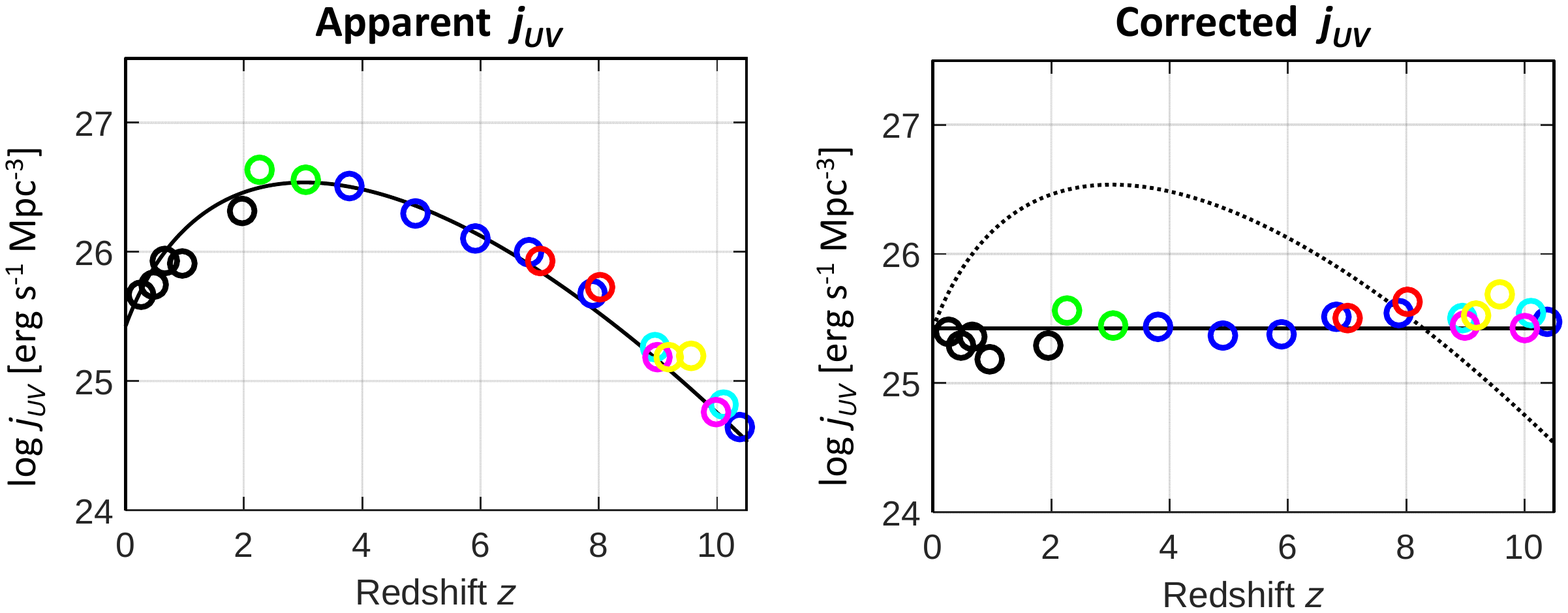}
\caption{
(a) The apparent comoving UV luminosity density as a function of redshift assuming a transparent universe. (b) The corrected UV luminosity density as a function of redshift in a dusty universe. Observations are taken from \citet[black circles]{Schiminovich2005}, \citet[green circles]{Reddy2009}, \citet[blue circles]{Bouwens2014a}, \citet[red circles]{McLure2013}, \citet[magenta circles]{Ellis2013}, \citet[cyan circles]{Oesch2014}, and \citet[yellow circles]{Bouwens2014b}. The solid line in (a) and the dotted line in (b) show the predicted proper luminosity density for the opaque universe with the UV intergalactic extinction of $0.075 \,\mathrm{mag}\, h\, \mathrm{Gpc}^{-1}$. The solid line in (b) shows the predicted comoving luminosity density corrected for the opaque universe with the UV intergalactic extinction of $0.075 \,\mathrm{mag}\, h\, \mathrm{Gpc}^{-1}$. 
}
\label{fig:11}
\end{figure*}

As shown in \citet{Vavrycuk2017a}, if the UV luminosity density is corrected for intergalactic opacity and transformed from the proper to comoving volumes according to equation (43), it becomes redshift independent, see Figure~\ref{fig:11}. The abundance of the apparent luminosity density at redshifts $2 < z < 4$ is commonly interpreted as the result of an enormously high SFR in this epoch \citep{Madau1998, Kochanek2001, Franceschini2008}. However, as indicated in Figure~\ref{fig:11}a, the luminosity density abundance at $2 < z < 4$ is actually caused by the expansion of the Universe. When going back in time, the Universe occupied a smaller volume and the proper number density of galaxies and the proper luminosity density produced by galaxies were higher (see Appendix A). The steep increase of the luminosity density is almost unaffected by intergalactic opacity for $z < 2$ because the Universe is effectively transparent at this epoch. Since the opacity of the Universe steeply increases with redshift and light extinction becomes significant for $z > 2-3$, the luminosity density does not increase further and starts to decline with $z$. This decline continues to very high redshifts. After correcting the luminosity density for intergalactic extinction, no decline at high redshifts is observed (Figure~\ref{fig:11}b).

\begin{figure*}
\includegraphics[angle=0,width=16.5cm,trim=60 140 80 140, clip=true]{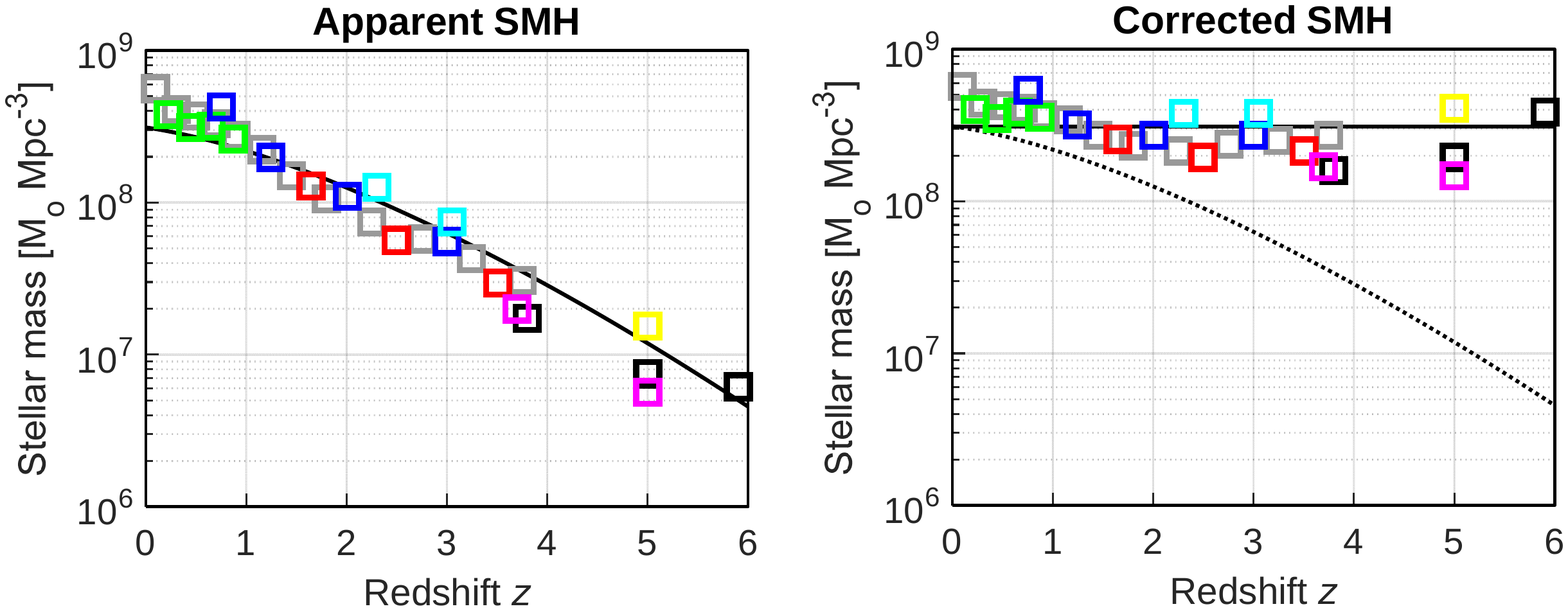}
\caption{
(a) The apparent global stellar mass history (Apparent SMH). The colour squares show observations reported by \citet[grey]{Perez_Gonzalez2008}, \citet[green]{Pozzetti2010}, \citet[blue]{Kajisawa2009}, \citet[red]{Marchesini2009}, \citet[cyan]{Reddy2012}, \citet[black]{Gonzalez2011}, \citet[magenta]{Lee2012}, and \citet[yellow]{Yabe2009}. The values are summarized in Table 2 of \citet{Madau_Dickinson2014}. The black line shows the apparent stellar mass history calculated using equation (26) of \citet{Vavrycuk2017a} with the UV intergalactic opacity of $0.08 \, \mathrm{mag} \, h \, \mathrm{Gpc}^{-1}$ at $z=0$. (b) The corrected global stellar mass history (Corrected SMH). The black dotted line shows the stellar mass history for a transparent universe, the black solid line shows the stellar mass history after eliminating the effect of the intergalactic opacity assuming $A_{UV}$ of $0.075 \, \mathrm{mag} \, h \, \mathrm{Gpc}^{-1}$ at $z=0$.
}
\label{fig:12}
\end{figure*}

The theoretical predictions of the corrected SMH calculated for an intergalactic opacity of $0.075 \, h \, \mathrm{mag} \, \mathrm{Gpc}^{-1}$ at UV wavelengths according to equation (41) are shown in Figure~\ref{fig:12}. If the transparent universe characterized by zero attenuation is assumed, observations suggest a steep decline of the stellar mass with redshift (Figure~\ref{fig:12}a). However, if intergalactic opacity is taken into account, the true SMH is constant and independent of redshift (Figure~\ref{fig:12}b). Hence the decline of the stellar mass density reported under the assumption of a transparent universe might be fully an artefact of neglecting the opacity of the intergalactic space. The redshift-independent comoving SMH (Figure~\ref{fig:12}b) looks apparently in contradiction with permanent star-formation processes in the Universe. However, it is physically conceivable provided the cosmic star formation rate is balanced by the stellar mass-loss rate due to, for example, core-collapse supernova explosions and stellar winds or superwinds \citep{Heckman2000, Woosley2002, Heger2003, Yoon2010}.

\section{Observations of submillimetre galaxies}

\begin{figure}
\includegraphics[angle=0,width=10cm,trim=90 40 10 40,clip=true]{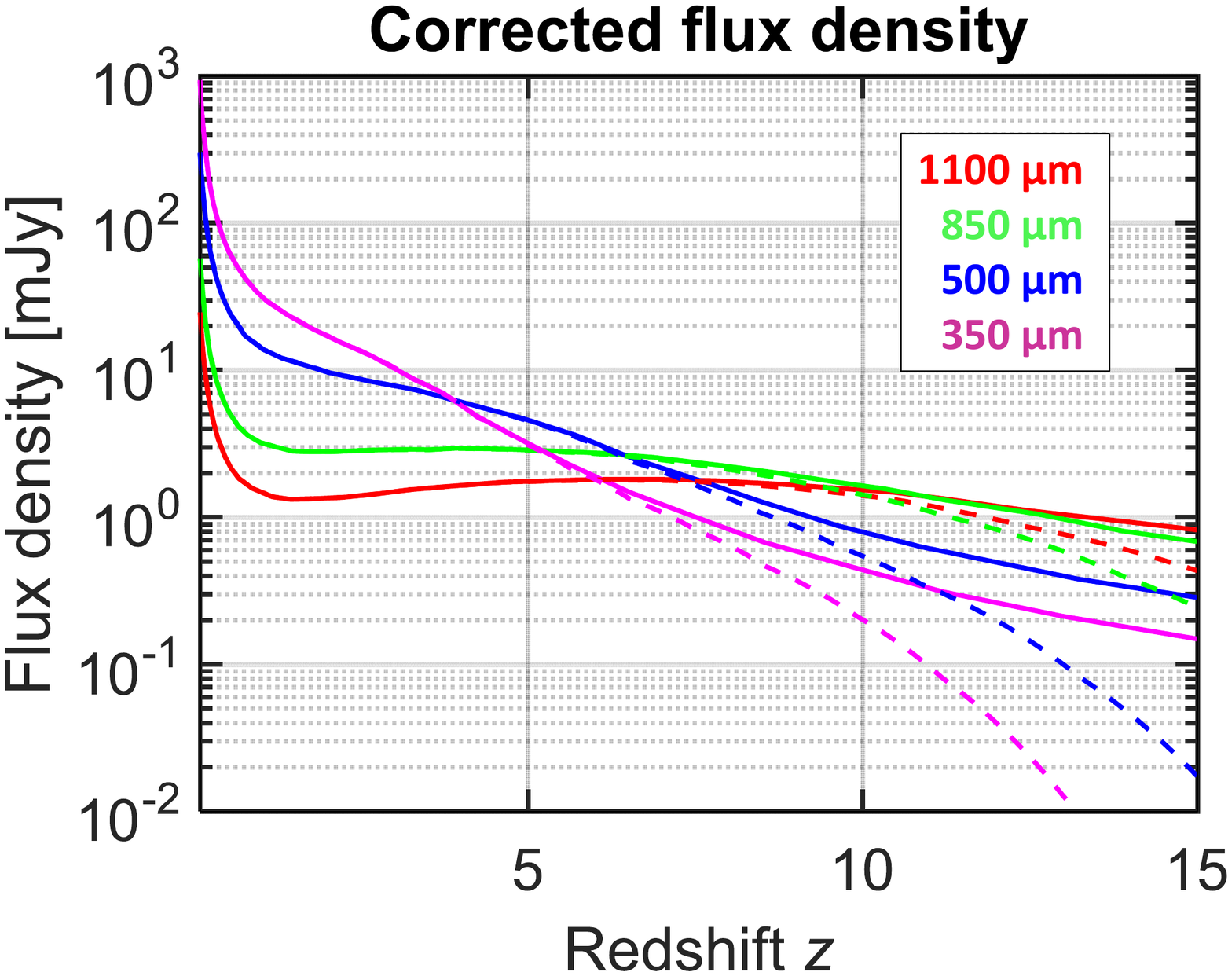}
\caption{
The predicted flux density of a dusty galaxy as a function of redshift at several observed submm wavelengths. Full line - predictions of \citet[their Figure 4]{Blain2002} for a transparent Universe for wavelengths of 1100 $\mu$m (red), 850 $\mu$m (green), 500 $\mu$m (blue) and 350 $\mu$m (magenta). Dashed line - a correction for a non-zero intergalactic opacity. Note the high K correction at wavelengths 850 $\mu$m and 1100 $\mu$m, which yields a flux density almost independent of redshift. For higher frequencies, this property is lost. The template spectrum is chosen to reproduce typical properties of dusty galaxies \citep[their Figure 2]{Blain2002}. The corrected curves were calculated according to \citet[his equation 7]{Vavrycuk2017b}. 
}
\label{fig:13}
\end{figure}

A promising tool for probing the evolution of the Universe at high redshifts are observations at submillimetre (submm) wavelengths using instruments such as SCUBA \citep{Holland1999}, MAMBO \citep{Kreysa1999}, SPIRE \citep{Griffin2010} and SCUBA-2 \citep{Holland2013}. The key points that make the submm observations attractive are: (1) their minimum distortion due to intergalactic attenuation, and (2) their ability to sample the spectral energy density (SED) of galaxies at wavelengths close to its rest-frame maximum of $ \sim 100 \mu\mathrm{m}$. As a consequence, large negative K-corrections should enable detecting galaxies at redshifts of up to $z \sim 20$ \citep{Blain2002, Casey2014}. However, the submm observations also have limitations. First, an accurate determination of redshifts is conditioned by following up the submm sources at other wavelengths which is often difficult. For example, matching submm sources to radio counterparts proved to be successful in identifying them at other wavelengths \citep{Chapman2005} but sources with no or very weak radio counterparts are lost in this approach. Second, the detection limit of $z \sim 20$ for submm galaxies is too optimistic because it neglects the frequency- and redshift-dependent intergalactic opacity. As shown in Figure~\ref{fig:13}, only a flux density at wavelengths greater than 1 mm is essentially undistorted for $z$ up to 15. The flux density at 350 or 500 $\mu\mathrm{m}$ starts to be markedly attenuated at redshifts higher than 10. Finally, the galaxy radiation targeted by the submm observations is the thermal radiation of galactic dust with a temperature of $\sim$30-40 K. Since the temperature of the intergalactic dust increases with redshift as $(1+z)\,T^{\mathrm{CMB}}$, the galactic and intergalactic dust have similar temperatures at redshifts of 10-15. Because of no or weak temperature contrast between the galaxies and intergalactic dust at these redshifts, the thermal radiation of galactic dust is lost in the background intergalactic radiation. Consequently, dusty star-forming galaxies with dust temperatures of 30-40 K cannot be observed at submm wavelengths at $z>10$.

As a result, so far only high-redshift galaxy samples of a limited size and restricted to redshifts of less than 5-6 are available \citep{Chapman2005, Coppin2009, Cox2011, Strandet2016, Ikarashi2017} which are not decisive enough for statistically relevant conclusions about the galaxy number density and properties of galaxies at redshifts $z>5$.

\section{Reionization and the Gunn-Peterson trough}

The high-redshift Universe can be studied by an evolution of the neutral hydrogen fraction in the IGM traced by observations of the Lyman $\alpha$ (Ly$\alpha$) forest of absorption lines in quasar optical spectra. These absorption lines are produced by neutral hydrogen (H I) in intergalactic gaseous clouds ionized by ultraviolet radiation at a wavelength of 1216 \AA  \citep{Wolfe2005, Meiksin2009}. The incidence rate of the absorption lines per unit redshift carries information about the number density of the Ly$\alpha$ clouds and its evolution in time and the width of the absorption lines depends on the column density of the clouds. 

According to the Big Bang theory, the neutral hydrogen in the IGM is reionized at redshifts between 6 and 20 by the first quasars and galaxies \citep{Gnedin_Ostriker1997, Gnedin2000b}, so that the IGM becomes transparent for the UV radiation at lower redshifts. Based on the interpretation of the CMB polarization as a product of the Thomson scattering \citep{Hu_White1997, Hu_Dodelson2002}, the reionization as a sudden event is predicted at $z \sim 11$ \citep{Dunkley2009, Jarosik2011}. Observations of the Ly$\alpha$ forest and the Gunn-Peterson trough in quasar spectra date the end of the reionization at $z \sim 6-7$ \citep{Becker2001, Fan2002, Fan2004, Fan2006b}. Obviously, if the Big Bang theory is not valid and the CMB has another origin, the idea of dark ages with mostly neutral hydrogen in the IGM and no sources of ionizing radiation is disputed. Similarly, the hypothesis about the reionization as an epoch of a transition from neutral to ionized hydrogen due to high-redshift galaxies and quasars is questioned.

Although, a change in the neutral hydrogen fraction by several tens of per cent at $z \sim 6-7$, supporting the idea of reionization, has been suggested by \citet{Pentericci2011}, \citet{Ono2012} and \citet{Schenker2012}, a rapid increase of neutral hydrogen is in conflict with a modelling of the evolution of the IGM, which favours the reionization as a rather gradual process with a continuous rise of the Ly$\alpha$ photons mean free path \citep{Miralda-Escude2000, Becker2007, Bolton_Haehnelt2013}. Moreover, it is unclear how sources with a declining comoving luminosity density at $z > 6$ (Figure~\ref{fig:11}a) could reionize the neutral hydrogen in the IGM \citep{Bunker2010, McLure2010, Grazian2011}. Instead, the Ly$\alpha$ optical depth measurements are more consistent with an essentially constant and redshift-independent photoionization rate \citep[their Figure 14]{Faucher_Giguere2008} predicted in Figure~\ref{fig:11}b and with no strong evolution in the neutral hydrogen fraction of the IGM \citep{Krug2012, Bolton_Haehnelt2013}. 

The redshift-independent ionization background radiation might also explain a puzzling ubiquity of the Ly$\alpha$ emission of very high-redshift galaxies if the IGM is considered to be significantly neutral over $7 < z < 9$ \citep{Stark2017, Bagley2017}. Hence, the evolution of the Ly$\alpha$ forest with redshift and the Gunn-Peterson trough at $z \sim 6-7$ might not be produced by increasing the abundance of neutral hydrogen in the Universe at $z > 6$ but by the Ly$\alpha$ clouds with a constant comoving neutral hydrogen fraction in a smaller proper volume of the Universe at high redshift. Since overdense Ly$\alpha$ regions with non-evolving neutral hydrogen fractions are close to each other, they start to touch and prevent escaping the Ly$\alpha$ photons.

\section{Evolution of metallicity}

Another tool for probing the history of the Universe is tracing heavy elements (metals) of the IGM with redshift. Models based on the Big Bang theory predict a persistent increase of the mean metallicity of the Universe with cosmic time driven mainly by star formation history \citep{Pei_Fall1995, Madau1998}. The mean metallicity should rise from zero to 0.001 solar by $z = 6$ (1 Gyr after Big Bang), reaching about 0.01 solar at $z = 2.5$ and 0.09 solar at $z = 0$ \citep[their Figure 14]{Madau_Dickinson2014}. Similarly to the mean metallicity of the Universe, the abundance of metals dispersed into the IGM by supernovae ejecta and winds is also continuously rising in the standard cosmological models \citep[their Figure 5]{Gnedin_Ostriker1997}. This can be tested by comparing the column density of a singly ionized metal to that of neutral hydrogen in Ly$\alpha$ systems. Frequently, DLA systems are selected because the ionization corrections are assumed to be negligible due to the high column density N(H I) producing the self-shielding of the DLA absorbers \citep{Prochaska2000, Vladilo2001}. 

However, observations do not provide convincing evidence of the predicted metallicity evolution. The observations indicate \citep{Rauch1998, Pettini2004, Meiksin2009}: (1) a puzzling widespread metal pollution of the IGM and a failure to detect a pristine material with no metals even at high redshifts, and (2) an unclear evolution of the metallicity. \citet{Prochaska2000} found no evolution in the N(H I)-weighted mean [Fe/H] metallicity for redshifts $z$ from 1.5 to 4.5, but later studies of larger datasets of DLAs have indicated a decrease of metallicity with increasing redshift \citep[ their Figure 1]{Prochaska2003}. \citet{Rafelski2012} combined 241 abundances of various metals (O I, S II, Si II, Zn II, Fe II and others) obtained from their data and from the literature, and found a metallicity decrease of -0.22 dex per unit redshift for $z$ from 0.09 to 5.06. The decrease is, however, significantly slower than the prediction and the scatter of measurements is quite large (up to 2 dex) making the result unconvincing \citep[their Figures 5 and 6]{Rafelski2012}. Furthermore, observations of the C IV absorbers do not show any visible redshift evolution over cosmic times from 1 to 4.5 Gyr after the Big Bang suggesting that a large fraction of intergalactic metals may already have been in place at $z > 6$ \citep{Songaila2001, Pettini2003, Ryan-Weber2006}.

\section{Primordial deuterium, helium and lithium abundances}

If the CMB is a thermal radiation of dust but not a relic radiation of the Big Bang, the concept of the Big Bang is seriously disputed. Firstly, except for the CMB, no direct observations indicate the Big Bang and no measurements provide information on the actual expanding/contracting history of the Universe at $z > 8-10$. Secondly, predictions of some cosmological constants and quantities based on the interpretation of the CMB anisotropies such as the baryonic density, helium abundance and dark matter density in the Universe, or timing of the reionization epoch at $z \sim 11$ \citep{Spergel2003, Spergel2007, Dunkley2009, Ade2015} are invalidated.

\begin{figure}
\includegraphics[angle=0,width=9cm,trim=70 60 10 40,clip=true]{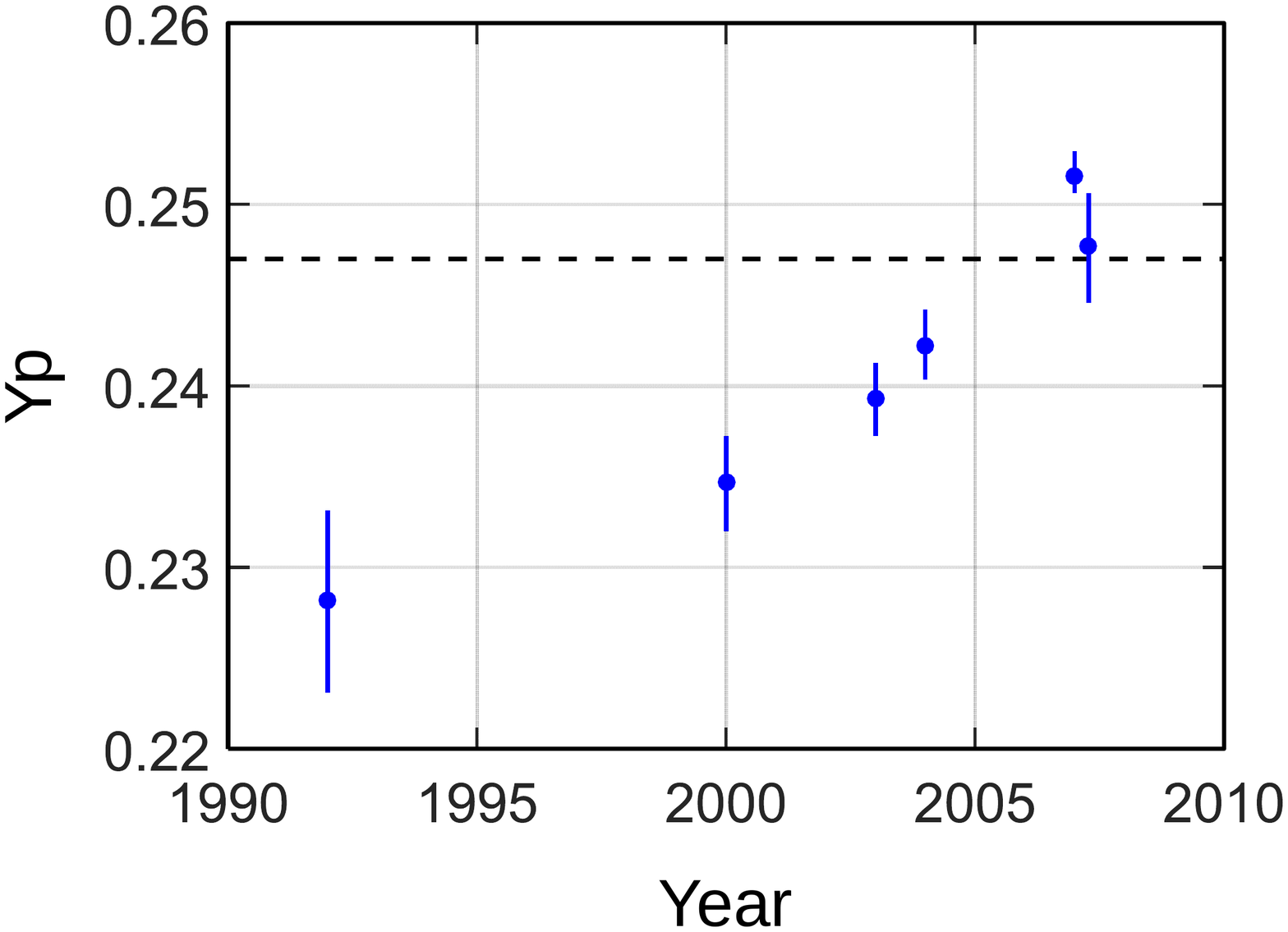}
\caption{
Measurements of the primordial helium abundance $Yp$ derived from H II regions with a small fraction of heavy metals. The measurements are taken from \citet{Pagel1992}, \citet{Peimbert2000}, \citet{Luridiana2003}, \citet{Izotov_Thuan2004}, \citet{Izotov2007} and \citet{Peimbert2007}. The values are summarized in \citet[his Table 1]{Peimbert2008}. The dashed line shows the theoretical prediction. Modified after \citet{Peimbert2008}.
}
\label{fig:14}
\end{figure}

The only remaining argument for the Big Bang theory is its prediction of primordial abundances of deuterium, helium and lithium in the Universe \citep{Olive2000, Cyburt2016}. The Big Bang nucleosynthesis (BBN) parameterizes the D, $^4$He and $^7$Li abundances by the baryon-to-photon ratio $\eta$ or equivalently the baryon density $\Omega_b \, h^2$. The baryon density is usually determined from deuterium abundance observations, so that the theory is capable of predicting only the other two values: the helium abundance $^4\mathrm{He/H} = 0.2470$ and the lithium abundance $^7\mathrm{Li/H} = 4.648 \times 10^{-10}$. Initially, observations did not match the predicted  $^4$He/H abundance well \citep{Pagel1992, Peimbert2000} but after two decades of efforts \citep{Peimbert2007, Izotov2014, Aver2015} when adopting a large number of systematic and statistical corrections \citep[their Table 7]{Peimbert2007}, a satisfactory fit has finally been achieved (Figure~\ref{fig:14}). By contrast, the fit of the lithium abundance is much worse; the predicted $^7$Li/H abundance is 2-3 times larger than observations \citep{Cyburt2008, Fields2011}. As stated by \citet{Cyburt2016}, to date, there is no solution of the discrepancy of the $^7$Li abundance without substantial departures of the BBN theory. Hence, the BBN theory may not provide us with fully-established firm evidence of the Big Bang.

\section{Dust theory and cyclic cosmology}

The model of a dusty universe is based on completely different postulates than the Big Bang theory. It is assumed that that global stellar mass density and the overall dust masses within galaxies and in intergalactic space are essentially constant with cosmic time. Consequently, the cosmic star formation rate is balanced by the stellar mass-loss rate due to, for example, core-collapse supernova explosions and stellar winds or superwinds. These constraints are needed for the EBL to rise as $(1+z)^4$ and the dust temperature to increase exactly as $(1+z)$ with redshift. These assumptions seem apparently unphysical and contradicting observations but most of these arguments are not actually well established and can be disproved as shown in the previous sections. 

If the number density of galaxies and the overall dust masses within galaxies and in the intergalactic space are basically constant with cosmic time, then this might happen within an oscillating model of the universe with repeating expansion and contraction periods. The cyclic cosmological model was originally proposed by Friedmann in 1922 \citep{Friedmann1999} and developed later in many modifications \citep{Steinhardt_Turok2002, Novello_Bergliaffa2008, Battefeld_Peter2015} including the quasi-steady-state cosmological model \citep{Narlikar2007}. Obviously, the idea of the universe oscillating within a given range of redshifts is a mere hypothesis full of open questions. The primary question is: which forces drive the oscillations. Without proposing any solution, we can just speculate that formations/destructions of galaxies and complex recycling processes in galaxies and in the IGM might play a central role in this model \citep{Segers2016, Angles-Alcazar2017}. Importantly, the upper limit of the redshift should not be very high $(z \lesssim 20-30)$ to allow the existence of galaxies as independent units even in the epoch of the minimum proper volume of the Universe. If so, the oscillations around a stationary state reflect only some imbalance in the Universe and the CMB comes partly from the previous cosmic cycle or cycles.

\section{Conclusions}

The analysis of the EBL and its extinction caused by opacity of the Universe indicates that the CMB might be thermal radiation of intergalactic dust. Even though the local Universe is virtually transparent with an opacity of $\sim 0.01 \, \mathrm{mag}\, h\, \mathrm{Gpc}^{-1}$ at visual wavelengths, it might be considerably opaque at high redshifts. For example, the visual opacity predicted by the proposed model reaches values of about 0.08, 0.19, 0.34 and 0.69 mag at $z$ = 1, 2, 3 and 5, respectively. Such opacity is hardly to be detected in the Type Ia supernova data \citep{Jones2013, Rodney2015, deJaeger2017}, but it can be studied using quasar data \citep{Menard2010a,Xie2015}. Since the energy of light is absorbed by dust, it heats up the dust and produces its thermal radiation. The temperature of dust depends on the intensity of light surrounding dust particles. Within galaxies, the light intensity is high the dust being heated up to $20-40\,\mathrm{K}$ and emitting thermal radiation at IR and FIR wavelengths. The intensity of light in intergalactic space is much lower than within galaxies, hence the intergalactic dust is colder and emits radiation at microwave wavelengths. 

The actual intergalactic dust temperature depends on the balance of the energy radiated and absorbed by galaxies and dust. The EBL energy radiated by galaxies and absorbed by intergalactic dust is re-radiated in the form of the CMB. The CMB radiation is mostly absorbed back by intergalactic dust but also partly by the dust in galaxies. The intergalactic dust is warmed by the EBL and the warming process continues until intergalactic dust reaches energy equilibrium. This happens when the energy interchanged between galaxies and intergalactic dust is mutually compensated. Hence the EBL energy absorbed by dust equals the CMB energy absorbed by galaxies. 

The distribution of the CMB energy between intergalactic dust and galaxies is controlled by the opacity ratio calculated from the galactic and intergalactic opacities. The opacity ratio is frequency and redshift independent and controls the temperature of intergalactic dust. A high opacity ratio means that a high amount of the CMB is absorbed back by dust. Consequently, dust is warmed up by the EBL to high temperatures. A low opacity ratio means that a significant part of the CMB energy is absorbed by galaxies, hence the dust is warmed up by the EBL to rather low temperatures. The opacity ratio can also be estimated from the EBL and CMB intensities. The optimum value of the opacity ratio calculated from observations of the galactic and intergalactic opacities is 13.4 while that obtained from the EBL and CMB intensities is 12.5. The fit is excellent considering rather high uncertainties in observations of the EBL and in the galactic and intergalactic opacities. 

The thermal radiation of dust is redshift dependent similarly as the radiation of any other objects in the Universe. Since its intensity basically depends on the EBL, which increases with redshift as $\left(1+z\right)^4$, the CMB temperature increases with redshift as $\left(1+z\right)$. The temperature increase with $z$ exactly eliminates the frequency redshift of the dust thermal radiation. Consequently, the radiation of dust observed at all distances looks apparently as the radiation of the blackbody with a single temperature. This eliminates the common argument against the CMB as the thermal radiation of dust that the spectrum of dust radiation cannot be characterized by a single temperature because of redshift \citep[p. 289]{Peacock1999}. The redshift dependence of the EBL intensity, CMB temperature and CMB intensity are invariant to the cosmology and can be applied to models of the Universe with a complicated expanding/contracting history. 

The CMB is radiated at a broad range of redshifts with the maximum CMB intensity coming from redshifts of 25-40 provided that the hitherto assumed expansion history of the Universe is correct. If the expansion history is different, e.g., if the Universe is oscillating, then part of the CMB might come from previous cosmic cycles. This indicates that the observed CMB stems from an enormous space and a long epoch of the Universe. As a consequence, the observed CMB temperature and intensity must be quite stable with only very small variations with direction in the sky. These variations reflect the EBL fluctuations due to the presence of large-scale structures as clusters, superclusters and voids in the Universe. The predicted CMB variation calculated from estimates of the EBL fluctuations attains values of tens of $\mu$K well consistent with observations. The CMB polarization is produced by a polarized emission of needle-shaped conducting dust grains present in the IGM aligned by cosmic magnetic fields around large-scale structures in the Universe. The phenomenon is fully analogous to a polarized interstellar dust emission in the Galaxy which is observed at shorter wavelengths because the temperature of the interstellar dust is higher than that of the intergalactic dust. Since the temperature and polarization anisotropies of the CMB have a common origin - existence of clusters, superclusters and voids - both anisotropies are spatially correlated.

The intensity of the CMB exactly corresponds to the intensity radiated by the blackbody with the CMB temperature. This implies that the energy flux received at a unit area of intergalactic space is equal to the energy flux emitted by a unit area of intergalactic dust particles. This statement is basically a formulation of the Olbers' paradox applied to dust particles instead of to stars. Since the sky is fully covered by dust particles and distant background particles are obscured by foreground particles, the energy fluxes emitted and received by dust are equal. Consequently, the intensity of the CMB does not depend on the actual dust density in the local Universe and on the expanding/contracting history of the Universe.

Further development of the dust theory depends on more accurate measurements of the EBL, distribution of the galactic and intergalactic dust, and the opacity of galaxies and of the intergalactic space at high redshifts. More definitive evidence of the properties of the high-redshift Universe can be provided by the James Webb Space Telescope \citep{Gardner2006, Zackrisson2011}. This telescope can probe the galaxy populations and properties of the IGM at high redshift and check which cosmological model suits the observations better.

\appendix
\section{EBL and luminosity density}

The bolometric intensity of the EBL (in $\mathrm{W m}^{-2}\mathrm{sr}^{-1}$) is calculated as an integral of the redshift-dependent bolometric luminosity density of galaxies reduced by the attenuation-obscuration effect \citep[his equation 1]{Vavrycuk2017a}
\begin{equation}\label{eqA1}
I_0^{\mathrm{EBL}} =\frac{1}{4\pi} \int_0^{z_{\mathrm{max}}} \frac{j\left(z\right)}{\left(1+z\right)^2}  
\,e^{-\tau \left(z\right)}\, \frac{c}{H_0} \frac{dz}{E\left(z\right)} \,,
\end{equation}
where $I_0^{\mathrm{EBL}}$ is the EBL intensity at present $(z = 0)$, $j(z)$ is the proper bolometric luminosity density, $\tau(z)$ is the bolometric optical depth, $c$ is the speed of light, $H_0$ is the Hubble constant, and $E\left(z\right)$ is the dimensionless Hubble parameter
\begin{equation}\label{eqA2}
E\left(z\right) = \sqrt{\left(1+z\right)^2\left(1+\Omega_m z\right)-z\left(2+z\right)\Omega_{\Lambda}} \,,
\end{equation}
where $\Omega_m$ is the total matter density, and $\Omega_{\Lambda}$ is the dimensionless cosmological constant. Equation (A1) is approximate because the optical depth is averaged over the EBL spectrum; a more accurate formula should consider the optical depth $\tau \left(z\right)$ as a function of frequency. 

Taking into account that the proper luminosity density $j(z)$ in Equation (A1) depends on redshift as \citep[his equation 5]{Vavrycuk2017a}
\begin{equation}\label{eqA3}
j\left(z\right) = j_0\left(1+z\right)^4 \,\,,
\end{equation}
where $j_0$ is the comoving luminosity density, we get
\begin{equation}\label{eqA4}
I_0^{\mathrm{EBL}} =\frac{1}{4\pi} \int_0^{z_{\mathrm{max}}} j_0 \left(1+z\right)^2  
\,e^{-\tau \left(z\right)}\, \frac{c}{H_0} \frac{dz}{E\left(z\right)} \,.
\end{equation}

Bear in mind that $j(z)$ in Equation (A1) is the proper luminosity density but not the comoving luminosity density as commonly assumed, see \citet[their equation 9]{Dwek1998} or \citet[their equation 5]{Hauser2001}. Integrating the comoving luminosity density in Equation (A1) would lead to incorrect results because it ignores the fact that we observe the luminosity density from different epochs of the Universe. When considering the proper luminosity density in Equation (A1) we actually follow observations \citep{Franceschini2001, Lagache2005, Franceschini2008} and sum the individual contributions to the EBL at various redshifts. The proper luminosity density in the EBL integral is used also by \citet{Peacock1999}. His formula is, however, different from Equation (A4) because he uses the reference luminosity density $j_0$ at early cosmic times. Obviously, fixing $j_0$ to the early cosmic times is possible and mathematically correct \citep[his equation 3.95]{Peacock1999} but not applicable to calculating the EBL using the luminosity density measured at $z = 0$.

Equation (A4) can also be derived from equation (A1) in an alternative straightforward way. The quantity $j\left(z\right)$ in equation (A1) means the bolometric luminosity density at $z$ and the factor $(1+z)^{-2}$ reflects reducing this luminosity density due to the expansion of the Universe. However, if we calculate the EBL from observations, we do not fix $j$ at redshift $z$ but at the present epoch ($z=0$). Hence, we go back in time and correct the luminosity density $j_0$ at $z=0$ not for the expansion but for the contraction of the Universe in its early times. As a consequence, instead of the factor $(1+z)^{-2}$ in equation (A1) we use the factor $(1+z)^2$ in equation (A4).

\section*{Acknowledgements} 
I thank an anonymous reviewer for helpful comments and Alberto Dom\'{i}nguez for providing me kindly with Figure 1. This research has made use of the NASA/IPAC Infrared Science Archive, which is operated by the Jet Propulsion Laboratory, California Institute of Technology, under contract with the National Aeronautics and Space Administration.


\bibliographystyle{mnras}

\bibliography{paper} 

\end{document}